\documentclass[preprint]{elsarticle}

\usepackage{geometry}
\usepackage{array}
\usepackage{siunitx}
\usepackage{amsmath}
\usepackage{cleveref}
\usepackage{graphicx}
\usepackage{subcaption}
\usepackage{xcolor}
\usepackage{mathrsfs}
\usepackage{ulem}
\usepackage{float}
\usepackage{makecell}

\biboptions{sort&compress}
\usepackage[version=4]{mhchem}

\Crefname{figure}{Fig.}{Figs.}
\Crefname{equation}{Eq.}{Eqs.}

\newcommand{\V}{{\text{V}_{\text{O}}^{..}}}
\newcommand{\VOne}{{\text{V}_{\text{O}}^{.}}}
\newcommand{\AccOne}{{\text{Acc}'}}
\newcommand{\AccTwo}{{\text{Acc}''}}
\newcommand{\AccZero}{{\text{Acc}^\times}}
\newcommand{\E}{{\text{e}'}}
\newcommand{\Hole}{{\text{h}^{.}}}

\newcommand{\subV}{_{\text{V}_\text{O}^{..}}}
\newcommand{\subVOne}{_{\text{V}_\text{O}^{.}}}
\newcommand{\subVb}{_{\text{V}_\text{O}^{..},\text{b}}}
\newcommand{\subVOneb}{_{\text{V}_\text{O}^{.},\text{b}}}
\newcommand{\subVc}{_{\text{V}_\text{O}^{..},\text{c}}}
\newcommand{\subVOnec}{_{\text{V}_\text{O}^{.},\text{c}}}
\newcommand{\subAccOne}{_{\text{Acc}'}}
\newcommand{\subAccOneb}{_{\text{Acc}',\text{b}}}
\newcommand{\subAccOnec}{_{\text{Acc}',\text{c}}}
\newcommand{\subAccTwo}{_{\text{Acc}''}}
\newcommand{\subAccTwob}{_{\text{Acc}'',\text{b}}}
\newcommand{\subAccTwoc}{_{\text{Acc}'',\text{c}}}
\newcommand{\subAccZero}{_{\text{Acc}^{\times}}}
\newcommand{\subAccZerob}{_{\text{Acc}^{\times},\text{b}}}
\newcommand{\subAccZeroc}{_{\text{Acc}^{\times},\text{c}}}
\newcommand{\subE}{_{\text{e}'}}
\newcommand{\subEb}{_{\text{e}',\text{b}}}
\newcommand{\subEc}{_{\text{e}',\text{c}}}
\newcommand{\subHole}{_{\text{h}^{.}}}
\newcommand{\subHolec}{_{\text{h}^{.},\text{c}}}
\newcommand{\subHoleb}{_{\text{h}^{.},\text{b}}}
\newcommand{\subAcc}{_{\text{Acc}}}
\newcommand{\subAccb}{_{\text{Acc},\text{b}}}
\newcommand{\subAccc}{_{\text{Acc},\text{c}}}
\newcommand{\bc}{k_\text{B}}

\newcommand{\be}{\begin{equation}}
\newcommand{\ee}{\end{equation}}

\newcommand{\pO}{P_{\text{O}_2}}

\begin{document}
\begin{frontmatter}



\title{Impact of charge transition levels on grain boundary properties in acceptor doped oxide ceramics: A phase-field study} 

\author[a,b]{Kai Wang\corref{cor1}}
\author[c,d,e]{Sangjun Kang}
\author[f]{Mahmoud Serour}
\author[g]{Roger A. De Souza}
\author[h]{Andreas Klein}
\author[i]{Rotraut Merkle}
\author[j]{Wolfgang Rheinheimer}
\author[c,d,e]{Christian K\"ubel}
\author[k]{Lijun Zhang}
\author[f]{Karsten Albe}
\author[a]{Bai-Xiang Xu}

\cortext[cor1]{Corresponding author}

\address[a]{Mechanics of Functional Materials Division, Institute of Materials Science, Technische Universit\"at Darmstadt, Darmstadt, 64287, Germany}
\address[b]{IMDEA Materials, Getafe, 28906, Madrid, Spain}
\address[c]{In situ Electron Microscopy, Department of Materials Science, Technische Universit\"at Darmstadt, 64287 Darmstadt, Germany}
\address[d]{Institute of Nanotechnology (INT), Karlsruhe Institute of Technology, 76344 Eggenstein-Leopoldshafen, Germany}
\address[e]{Karlsruhe Nano Micro Facility (KNMFi), Karlsruhe Institute of Technology (KIT), 76344 Eggenstein-Leopoldshafen, Germany}
\address[f]{Materials Modelling Division, Institute of Materials Science, Technische Universit\"at Darmstadt, Darmstadt, 64287, Germany}
\address[g]{Institute of Physical Chemistry, RWTH Aachen University, Aachen, 52056, Germany}
\address[h]{Institute of Materials Science, Electronic Structure of Materials, Technische Universit\"at Darmstadt, 64287 Darmstadt, Germany}
\address[i]{Max Planck Institute for Solid State Research, Heisenbergstra\ss e 1, Stuttgart, 70569, Germany}
\address[j]{University of Stuttgart, Institute for Manufacturing Technologies of Ceramic Components and Composites (IFKB), Stuttgart, 70569, Germany}
\address[k]{State Key Laboratory of Powder Metallurgy, Central South University, Changsha 410083, China}


\begin{abstract}
Advanced doping strategies enable the functionalities of oxide ceramics by tailoring bulk defect chemistry and space charge layer (SCL) behavior at interfaces. Charge transition levels (CTLs), defined as the Fermi level at which a defect changes its most stable charge state, are central to this process. Their alignment with the Fermi level governs bulk defect chemistry, while their bending within the SCL induces additional charge-state transitions. Thus, incorporating CTLs is essential for a consistent description of defect equilibria and SCL formation in systems with accessible CTLs.
In this work, we propose for the first time a defect-chemistry consistent phase-field model explicitly coupled with CTLs to investigate their role in SCL evolution.
The model includes multiple defect species, including multivalent oxygen vacancies, multivalent acceptor dopants, electrons, and holes. 
This model is applied to Fe-doped SrTiO$_3$ for wide ranges of oxygen partial pressures and temperatures, capturing both symmetric SCL profiles at stationary grain boundaries and asymmetric SCL formation during migrating ones. In particular, two distinct grain boundary types, slow and fast boundaries, emerge during migration, supported by transmission electron microscopy observations. The simulations reveal that CTL-governed bulk defect chemistry, together with CTL bending induced additional charge-state transitions within SCL, critically determine SCL characteristics.  Moreover, CTL-mediated hole transport is significantly faster than the diffusion of acceptor dopants, thereby modulating solute drag strength and grain boundary kinetics. Finally, this model is applied to predict the grain boundary properties that depend simultaneously on thermal history and grain boundary type. Simulation results reveal that slow and fast grain boundaries, induced by solute drag effects, exhibit distinctly different property values. 
This model provides a unified framework for linking defect chemistry, Fermi level, CTLs, and grain boundary kinetics and properties, offering new insights into the design of oxide ceramics with tailored functional properties.


\end{abstract}





\end{frontmatter}

\section{Introduction} \label{introduction}

Oxide ceramics, based on SrTiO$_3$ (STO) and BaTiO$_3$ (BTO), are widely used in a broad range of functional devices, including capacitors \cite{hou2017ultrahigh,stengel2006origin,tang2013synthesis}, actuators \cite{kursumovic2013new, gao2017recent}, sensors \cite{wang2022stretchable, chan2014highly}, memristors \cite{muenstermann2010coexistence, molinari2020configurable}, and electrolytes\cite{guo2025cobalt,shah2020semiconductor}, owing to their versatile electrical, electrochemical and electromechanical properties.  To tailor and enhance the performance of these materials for specific applications, advanced doping strategies have attracted increasing attention \cite{feng2020defects,xiong2024doping}. The introduction of dopant elements into the host lattice affects defect chemical equilibria, resulting in altered charged point defects whose concentrations and charge states are highly sensitive to temperature, oxygen partial pressure, and the electronic structure of the material \cite{maier1993defect}. In acceptor-doped oxide ceramics, oxygen vacancies, acceptor dopants, electrons, and holes are typically the primary charged species. The interplay among these charged species complicates the compensation landscape and poses a significant challenge for predictive modeling frameworks that aim to describe defect segregation, defect transport, and in particular, microstructural evolution.

Among these charged defects, the ionization behavior of acceptor dopants plays a particularly critical role in determining the charge compensation mechanism \cite{maier2023physical}. Unlike fully ionized species, many acceptor dopants can exist in multiple valence states under varying thermodynamic conditions \cite{klein2023fermi}. Transitions between different charge states, such as from neutral to singly ionized or from singly to doubly ionized, are governed by well-defined energy thresholds known as charge transition levels (CTLs). Each CTL corresponds to the Fermi level at which the defect changes its most favorable thermodynamically stable charge state. 
Consequently, the relative alignment between the Fermi level and the CTLs directly controls the degree of dopant ionization, thereby significantly affecting defect equilibria, charge compensation, and electrical response. The ability to tune Fermi levels is important for developing electronic device materials \cite{yang2014tuning}. Recently, the Fermi level has even been proposed as a common parameter for describing charge-compensation mechanisms in oxide ceramics (Fermi level Engineering) \cite{klein2023fermi}.

The CTLs of defects are material-specific, and even the same dopant can exhibit different compensation behavior in different host materials \cite{klein2023fermi}. CTLs of dopant are governed by the dopant’s electronic configuration, lattice site occupation, and the local bonding environment. X-ray photoelectron spectroscopy (XPS) can be applied to directly
determine CTLs of dopants in oxides \cite{chaoudhary2025direct, chaoudhary2025uncovering, liu2025co}.  
Fe-doped STO and BTO exemplify how CTLs vary across perovskite hosts and influence defect chemistry. In these systems, Fe substitutes on the Ti site and can adopt multiple oxidation states, Fe$^{4+}$, Fe$^{3+}$, and Fe$^{2+}$, depending on the position of the Fermi level relative to the CTLs. The Fe$^{{4+}/{3+}}$ CTL is located approximately 0.9~eV above the valance band maximum (VBM) in both STO and BTO, enabling Fe$^{3+}$ to act as a dominant acceptor state under moderately oxidizing conditions \cite{suzuki2019energy}. However, the Fe$^{{3+}/{2+}}$ CTL exhibits strong material dependence. In BTO, it appears at 2.4~eV above the VBM, thereby allowing stabilization of the Fe$^{2+}$ state under reducing conditions. In contrast, in STO, the same CTL lies at 2.9~eV above the VBM, making the Fe$^{2+}$ state energetically unfavorable across most accessible thermodynamic regimes \cite{suzuki2019energy}.  Many other transition-metal acceptors, such as Mn \cite{wechsler1988thermodynamic}, Cu \cite{suzuki2020fermi}, V \cite{chaoudhary2025direct} and Ni \cite{bonkowski2024single} also exhibit multiple CTLs within the band gap of various oxide hosts.

A rigorous description of defect concentrations in acceptor-doped oxide ceramics requires the explicit incorporation of defect chemistry and CTLs to accurately determine the thermodynamically stable charge states under given thermodynamic conditions. In the bulk, defect equilibria are governed by the position of the Fermi level relative to the CTLs, and are further constrained by mass conservation, local charge neutrality, and the laws of mass action \cite{wang2016defect, usler2024space, lohaus2023defect}.
In the vicinity of grain boundaries (GBs), the situation becomes more complex due to the presence of space charge layers (SCLs) induced by defect segregation to the GB core \cite{lohaus2023defect, de2009formation, de2019effect, mcintyre2000equilibrium}. In the case of titanate perovskites, oxygen vacancies segregate to the GB core, driven by a reduction in their formation energy relative to the bulk. This segregation creates a positively charged GB core, which is compensated by negatively charged species such as singly or doubly ionized acceptor dopants and electrons, thereby establishing local electrochemical equilibrium \cite{de2009formation, de2019effect}. 
The equilibrium SCL is numerically calculated in Mott–Schottky and Gouy–Chapman cases by using abrupt GB-SCL model \cite{jamnik1995interfaces}. In this model, the material parameters such as the standard formation energy of defects and the available sites of defects are assumed to change discontinuously across the GB core and the bulk. This approach has been successfully applied to Fe-doped STO with the consideration of Fe$^{3+/4+}$ transition and enables the prediction of key features of SCL formation at a stationary GB core, such as the space charge potential, the defect concentrations within the GB core, and the spatial distributions of both electrostatic potential and point defects in the adjacent SCL region \cite{de2009formation,usler2024space}.
In addition, the inhomogeneous electrostatic potential across the SCL gives rise to significant band bending, which shifts the local Fermi level relative to the band edges. Because the thermodynamic stability of defect charge states is determined by the relative position of the Fermi level with respect to the CTLs, this band bending induces a spatial variation in their alignment. As a result, dopants can undergo charge-state transitions across the SCL even without any change in the total dopant concentration. These transitions are driven purely by local electrostatics and can substantially alter the local defect distribution and charge-compensation mechanism.

Beyond the impact of CTL on defect equilibria at stationary GB, CTLs also dynamically modulate GB behavior during sintering and grain growth. In particular, charge-state transitions of dopants can also strongly influence the GB kinetics.  
SCLs introduce pronounced spatial inhomogeneities in the defect landscape. The segregation of acceptor dopants to the GB core leads to a substantial reduction in GB mobility through solute drag \cite{cahn1962impurity}, wherein the low diffusivity of dopants effectively “pins” the GB and hinders its migration during sintering. This interaction disrupts conventional grain growth kinetics, giving rise to non-Arrhenius behavior and abnormal grain growth (AGG) in systems such as Fe-doped STO \cite{rheinheimer2015non,rheinheimer2016grain}. While AGG has been attributed to either GB anisotropy \cite{rohrer2005influence,rohrer2011grain} or solute drag \cite{kim2008grain}, recent experimental evidence suggests that SCL-induced solute drag is the dominant mechanism in STO \cite{rheinheimer2016grain,zahler2023grain}. In addition, the co-existence of two types GBs, slow GB and fast GB, have also been observed as a consequence of solute drag effect in the numerical simulations \cite{wang2024defect,vikrant2020electrochemically}. Slow GBs are strongly affected by the solute drag effect due to the fact that the dopants remain segregated to the moving GBs, leading to nearly symmetric (or slightly asymmetric) profiles of both acceptor dopant concentration and electrostatic potential across the GB core. In contrast, fast GBs with less dopant segregation experience weaker solute drag effects, enabling faster boundary migration and resulting in strongly asymmetric dopant and potential profiles. {The existence of asymmetric SCLs and two types of GBs is also supported by our Scanning transmission electron microscopy (STEM) combined with energy-dispersive X-ray spectroscopy (EDS) observations, which will be described in detail in \Cref{Exp_section}.} 

Despite extensive research into solute drag phenomena in oxide ceramics, the role of CTLs has remained largely unexplored. Most existing numerical studies consider dopants as fixed-charge species, interacting electrostatically with the SCL \cite{vikrant2020electrochemical,vikrant2020electrochemically,wang2024defect}. However, dopant charge states are not fixed, and multiple charge states, including neutral ones, can become thermodynamically relevant. 
These neutral species interact more weakly with electrostatic fields and may contribute less effectively to solute drag, depending on their local charge state distribution.
Moreover, charge-state transitions are mediated by rapid electronic processes, such as hole or electron exchange, that are several orders of magnitude faster than cation diffusion. As a result, the charge-state distribution of dopants can re-equilibrate almost instantaneously in response to changes in GB position or local potential. This indicates that solute drag is not solely determined by dopant diffusion, but is also continuously modulated by the dynamic redistribution of dopant charge states across the SCL.
Furthermore, GB migration alters the electrostatic potential landscape, inducing band bending that shifts the local Fermi level relative to the CTLs. These shifts change the thermodynamic stability of different dopant charge states and lead to corresponding changes in their spatial distribution.
The resulting charge redistribution and potential profile are also influenced by the type of GB, as slow and fast boundaries exhibit distinct symmetries in their SCL structures. These symmetry differences further modify the local band bending, thereby affecting the alignment between CTLs and the Fermi level.
Taken together, these considerations suggest that CTLs are an essential, yet overlooked, parameter in understanding and modeling solute drag phenomena under different thermodynamic conditions. Incorporating CTLs may offer a more comprehensive framework for interpreting solute drag and GB kinetics during sintering in oxide ceramics.

To capture the complex interplay among CTL-governed dopant ionization behavior, defect chemistry consistent SCL formation and GB migration, a modeling framework is required that self-consistently couples electrostatics, defect chemistry, Fermi level, CTLs and microstructural evolution.
The phase-field method, based on a diffuse-interface formalism, has emerged as a powerful and flexible approach for quantitatively simulating the moving boundary problems \cite{wang2020modeling, wang2021quantitative} and such strongly coupled phenomena during sintering process \cite{yang20193d, wang2024defect, oyedeji2023variational}.
With respect to electrostatic interactions, pioneering work has been carried out by Guyer et al., who investigated the thermodynamic equilibrium and kinetic behavior of electrochemical interfaces within a phase-field framework \cite{guyer2004phaseEquilibrium,guyer2004phaseKinetics}.   Garc\'ia et al. proposed a thermodynamically consistent variational framework to model the time evolution of electrically and magnetically active materials, opening avenues for simulating complex coupled systems \cite{garcia2004thermodynamically}. 
Vikrant et al. developed a phase-field model based on the Kobayashi-Warren-Carter framework \cite{kobayashi2000continuum} for grain growth, and investigated the formation of SCLs under both equilibrium and quasi-equilibrium conditions \cite{vikrant2020electrochemical}. Their model explicitly incorporates the effects of GB misorientation and solute drag, providing insights into the AGG and bimodal grain size distribution in Fe-doped STO during sintering \cite{vikrant2020electrochemically}.
Aagesen et al. proposed a grand potential based phase-field model and included the effects of charged vacancies and the associated interactions between internal and applied electric fields \cite{aagesen2024electrochemical}.
Wang et al. have developed a defect-chemistry-informed phase-field framework \cite{wang2024defect}.
This model explicitly accounts for the available site densities of oxygen vacancies and singly ionized acceptor dopants in both the bulk phase and the GB core, enabling a consistent description of configurational entropy contributions to the electrochemical free energy.
The predicted SCL formation in Fe-doped STO agrees well with results obtained from the abrupt GB$|$SCL model \cite{de2009formation}. Furthermore, the model has been applied to investigate solute drag effects and their influence on skewed grain size distributions that deviate from the conventional log-normal behavior during sintering.

Despite significant progress in phase-field modeling of SCL formation and grain growth process in oxide ceramics, several important questions remain open.
First, beyond the commonly considered singly ionized state in previous studies \cite{vikrant2020electrochemical, vikrant2020electrochemically, wang2024defect}, it is not yet fully understood how different dopant charge states contribute to defect equilibria and electrostatic behavior under varying thermodynamic conditions.
Second, it remains unclear how charge-state transitions governed by the alignment between the Fermi level and CTLs can be consistently incorporated into phase-field formulations, particularly in capturing their impact on defect distributions and electrostatic potential profiles.
Third, while solute drag effects have been widely studied in the context of dopant segregation, it is still an open question how the thermodynamic competition between different dopant charge states influences the solute drag behavior. In particular, the role of rapid electronic processes in enabling dynamic re-equilibration of dopant charge states during GB migration, and its impact on solute drag strength, remains to be clarified.
Addressing these questions requires the extension of the phase-field model that explicitly accounts for CTL-governed charge-state transitions and their coupling with defect chemistry and GB kinetics.


Therefore, we aim to develop a defect-chemistry-consistent phase-field model that explicitly incorporates Fermi level and CTLs. This enables a thermodynamically rigorous description of multivalent acceptor dopant ionization across a wide range of temperatures and oxygen partial pressures. The model captures the spatial and temporal redistribution among neutral, singly, and doubly ionized dopant states, and resolves their impact on SCL formation, charge compensation, and the local electronic structure under both equilibrium and quasi-equilibrium conditions. More importantly, the model accounts for rapid charge-state transitions mediated by hole transport. This coupling allows for the dynamic modulation of solute drag strength during GB migration, offering new insights into the role of CTLs and multivalent dopants in controlling grain growth kinetics.

Beyond its theoretical significance, the proposed phase-field model provides practical value for interpreting and predicting experimentally observed phenomena. A key application lies in predicting GB properties that depend on both thermal history and GB type, particularly after sintering and quenching. At high sintering temperatures, acceptor dopants can reach thermodynamic equilibrium distributions. Upon subsequent cooling to the lower temperatures relevant for electrical characterization, such as those used in electrochemical impedance spectroscopy (EIS), their redistribution becomes kinetically hindered and effectively frozen \cite{guo2003blocking, swaroop2005lattice, usler2024space}.
Conventional analytical solution interpreting EIS spectra often adopt the Mott–Schottky model, which assumes immobile acceptor dopants uniformly distributed throughout the SCL \cite{gregori2017ion}. This assumption, however, conflicts with experimental evidence. Advanced characterization methods, including transmission electron microscopy (TEM) and atom probe tomography (APT), consistently reveal dopant accumulation near GB cores \cite{tian2000ionic, lei2002segregation, li2011direct, diercks2016three, xu2020variability}. More recently, EIS measurements have revealed the coexistence of multiple GB types with distinct electrical properties \cite{zahler2025non}. Notably, different GB types experience different solute drag effects, leading to variations in the distribution of acceptor dopants as discussed above. Together, these findings demonstrate that GB behavior is strongly influenced by both thermal history and GB type. Although the present model is not designed for direct quantitative comparison with experimental data, it provides a valuable framework for capturing these dependencies and for rationalizing how thermal history and GB type collectively govern GB properties.

The outline of this paper is as follows.
Section 2 demonstrates the experimental observation of two types of GBs (symmetric and asymmetric ones).
Section 3 presents a defect-chemistry-consistent phase-field model that explicitly incorporates the Fermi level and CTLs, with application to Fe-doped STO. Section 4 describes the numerical implementation of the phase-field model using a finite difference scheme accelerated by graphics processing units (GPUs).
The simulation results are presented in Section 5. We first reproduce the symmetric SCL formation in bicrystalline Fe-doped STO under various thermodynamic conditions at stationary GBs. We then investigate the formation of slow and fast GBs and the influence of CTLs on solute drag behavior during GB migration. Finally, we apply the model to simulate SCL reformation in slow and fast GBs after quenching and compare their electrical properties.
Section 6 summarizes the key findings and discusses the implications of this phase-field model for acceptor-doped oxide ceramics.

\section{Experimental observation of two types of grain boundaries}
\label{Exp_section}

\begin{figure}[h]
    \centering
    \includegraphics[width=1\linewidth]{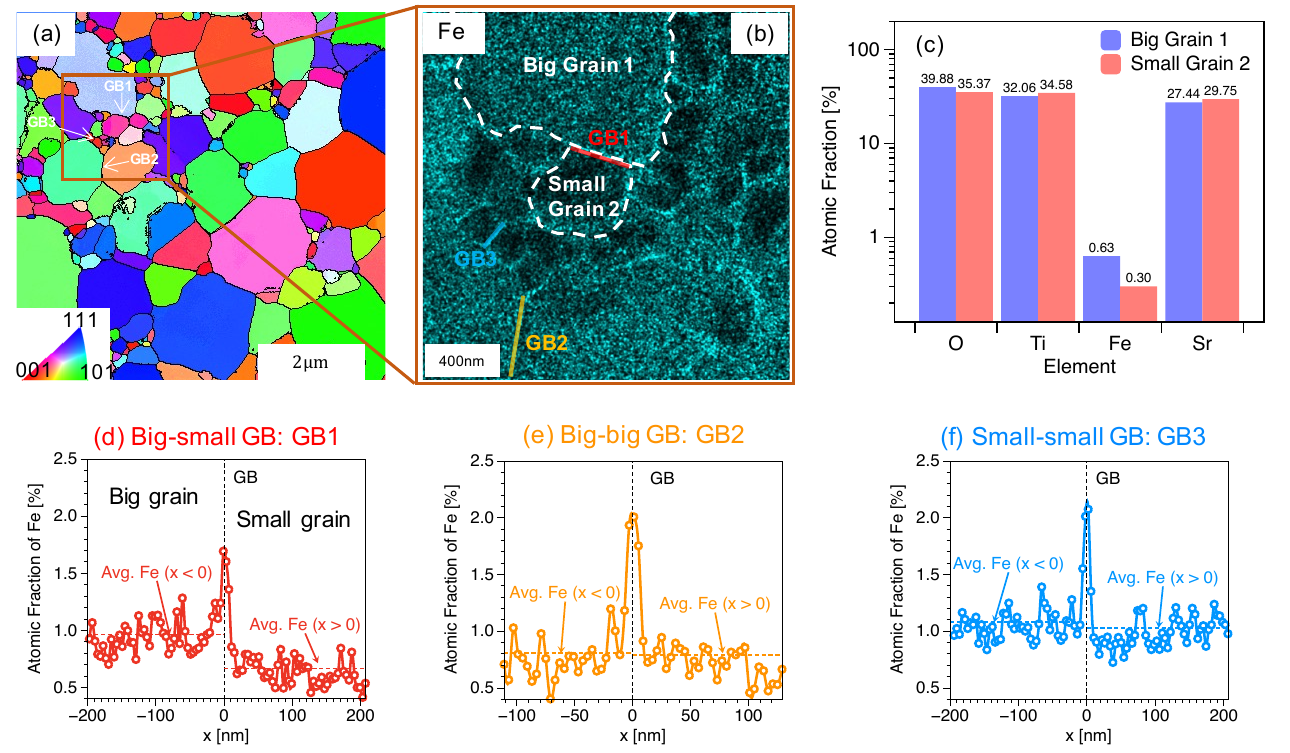}
   \caption{
STEM–EDS characterization of Fe segregation at representative GBs in a 2\% Fe-doped sample. 
(a) Crystallographic orientation map highlighting three types of GBs: large–small, large–large, and small–small. 
(b) STEM–EDS map of Fe distribution in the selected region. 
(c) Average atomic fractions of O, Ti, Fe, and Sr in large and small grains, showing comparable O, Ti, and Sr concentrations, while the Fe concentration in the small grain is approximately half that in the large grain. 
(d–f) Fe concentration profiles across the big–small, big–big, and small–small GBs, respectively. The big–small GB exhibits a pronounced asymmetric Fe distribution, whereas the big–big and small–small GBs show symmetric profiles.
}
    \label{Experiment}
\end{figure}

The formation of asymmetric SCLs at different GBs is supported by experimental observations, as shown in \Cref{Experiment}. Cross-sectional TEM lamellae were prepared from 2\% Fe-doped specimens. This relatively high dopant concentration was intentionally chosen to enhance the detectability of GB segregation and improve the signal-to-noise ratio in STEM–EDS measurements, while preserving the same qualitative defect chemistry and segregation mechanisms as in lower dopant concentrations.

Lamellae were prepared using focused ion beam (FIB) lift-out, followed by low-energy ion milling to minimize surface damage. GBs were identified in STEM mode and characterized using high-angle annular dark-field (HAADF) imaging and automated crystal orientation mapping (ACOM). Elemental distributions across selected GBs were analyzed by STEM–EDS spectrum imaging. To reduce channeling-related artifacts, the specimen was slightly tilted off-zone during EDS acquisition when necessary. The EDS data were processed using standard procedures, including background subtraction, peak deconvolution, and elemental mapping. Line profiles normal to the GB plane were extracted by integrating intensities over a narrow rectangular region. Due to the influence of absorption and channeling effects in thin foils, the EDS results are presented in terms of relative elemental enrichment or depletion across the GB, rather than absolute site fractions.

\Cref{Experiment}(a) shows the crystallographic orientation map of the sample, highlighting distinct orientations of neighboring grains. Three representative GBs (GB1, GB2, and GB3) are identified, corresponding to GBs between big–small, big–big, and small–small grains, respectively. \Cref{Experiment}(b) presents a representative Fe STEM–EDS map of the selected region, indicating the Fe distribution in both the bulk and along the GBs marked in \Cref{Experiment}(a). The positions of GB1, GB2, and GB3 are indicated by red, yellow, and blue solid lines, respectively. The compositional profiles of a representative large grain (grain 1) and small grain (grain 2), which are indicated by white dashed lines, and the corresponding elemental distributions are shown in \Cref{Experiment}(c).
During grain growth, the atomic fractions of O, Ti, and Sr in the bulk remain relatively uniform across grains. In contrast, the Fe concentration in the small grain is approximately half of that in the large grain. Since the small grain is shrinking during grain growth, this observation indicates Fe depletion in the shrinking grain.
\Cref{Experiment}(b)–(d) further present the Fe distributions across the three GB types. For three GBs, we observe the dopant segregation in the GB core. The segregated Fe concentrations in big-big and small-small GBs are higher than big-small GBs. For the big–small GB, a pronounced asymmetric Fe distribution is observed, with a higher Fe concentration on the side of the large grain. In contrast, for the big–big and small–small GBs, the Fe distributions are nearly symmetric across the boundaries.

The experimental observations demonstrate that different types of GBs can coexist within the same sample and influence the measured electrical response. The formation of these GB types is strongly governed by the thermodynamic conditions during sintering, such as temperature and oxygen partial pressure, which determine the Fermi level and defect equilibria.  In addition, CTLs determine the accessible charge states of dopants under these conditions, thereby affecting both the segregation behavior and SCL formation. To describe these coupled effects, a theoretical framework is required that explicitly accounts for thermodynamic conditions, CTL-governed defect chemistry, and the GB migration. In the following section, we develop a defect-chemistry consistent phase-field model to address these aspects.

\section{Phase-field grain growth model coupled with defect charge transition}

In our previous work, a defect-chemistry-informed phase-field model was proposed~\cite{wang2024defect}, extending existing approaches by explicitly accounting for different available site densities of defects in the bulk and in the GB core, as well as their coupling to electrostatics. This framework enabled accurate prediction of SCL formation in Fe-doped STO, consistent with defect chemistry theory.
Nevertheless, the charge state of Fe was restricted to the $3+$ state, and charge-state transitions were not considered. However, under varying thermodynamic conditions, charge transitions may occur and significantly influence SCL formation.
To address this limitation, the present work extends the previous model~\cite{wang2024defect} by explicitly incorporating multiple charged species, including multivalent acceptor dopants, oxygen vacancies, electrons, and holes. Importantly, incorporating multivalent dopants into the phase-field framework involves more than simply adding new concentration fields and evolution equations. It requires constructing a fully self-consistent set of governing equations, wherein the evolution of each species is constrained by defect chemistry and CTLs.
Indeed, the setup of our previous model~\cite{wang2024defect}, which introduced different available number densities of defects in the bulk and in the GB core, allow the extension to couple the defect charge transition in a defect-chemistry consistent manner, as detailed in the following.

The phase-field model here is dedicated to an acceptor-doped perovskite (ABO$_3$). The formation of multiple charge species are determined via the following defect reactions \cite{wang2016defect,usler2024space}. Reduction of the oxide produces oxygen vacancies and electrons:
\be \label{Rred}
\text{O}^{\times} \rightleftharpoons {\V} + \frac{1}{2}\text{O}_2 + 2\E.
\tag{$R\text{red}$}
\ee
If sufficient electrons are present, and if the temperature is sufficiently low, the oxygen vacancies may become singly ionized. The change in ionization is expressed by
\be \label{Rred1}
\V + \E \rightleftharpoons \VOne.
\tag{$R_{\text{V}_\text{O}1\rightarrow 2}$}
\ee
The neutral oxygen vacancies may also appear \cite{moos1997defect}. However, the transition level between singly ionized and neutral oxygen vacancies is only 3~meV below the conduction band minimum (CBM). Therefore, the influence of neutral oxygen vacancy is negligible and not considered for the temperature chosen in the present work. To compensate for the positively charged oxygen vacancies, B-site atoms in the lattice are partially substituted by acceptor dopants, which undergo successive ionization reactions to lower their valence states:
\be \label{Rion1}
\AccZero \rightleftharpoons \AccOne + \Hole,
\tag{$R_\text{ion1}$}
\ee
\be \label{Rion2}
\AccOne \rightleftharpoons \AccTwo + \Hole,
\tag{$R_\text{ion2}$}
\ee
In parallel, the generation and recombination of electrons ($\E$) and holes ($\Hole$) is governed by the intrinsic electronic reaction:
\be \label{Reh}
{nil} \rightleftharpoons \E + \Hole.
\tag{$R_\text{eh}$}
\ee
Here, $\V$, $\VOne$, $\AccZero$, $\AccOne$, $\AccTwo$, $\E$, and $\Hole$ represent the doubly ionized oxygen vacancy, singly ionized oxygen vacancy, neutral acceptor dopant, singly ionized acceptor dopant, doubly ionzied acceptor dopant, electron, and hole, respectively, in Kröger–Vink notation \cite{kroger1956relations}. $nil$ denotes the thermodynamic standard state of the crystal in which all of the electronic carriers are in their ground state. 
Based on these reactions, the thermodynamic equilibrium concentrations of all charged species are governed by mass action laws. Each reaction is associated with a corresponding mass reaction constant (e.g., $K_\text{red}$, $K_{\text{V}_\text{O}1\rightarrow 2}$, $K_\text{ion1}$, $K_\text{ion2}$, and $K_\text{eh}$). 
The CTLs associated with the $\AccZero/\AccOne$ and $\AccOne/\AccTwo$ transitions, denoted by $E_\text{A1}$ and $E_\text{A2}$ respectively, enter the expressions of $K_\text{ion1}$ and $K_\text{ion2}$. 

As such, the role of CTLs in the phase-field framework manifests in two key aspects. First, CTLs critically determine the initial defect concentration profiles by defining the equilibrium charge-state distribution of acceptor dopants under given thermodynamic conditions. Second, the interconversion among multivalent dopant species must be dynamically coupled to their ionization reaction, making CTLs a central component in the reaction kinetics, which in turn govern the time evolution of dopant concentrations. In the following, we provide a more detailed formulation of these effects within the phase-field framework.

\subsection{Free energy density functional}\label{Free energy density functional}

The free-energy-based Kim–Kim–Suzuki (KKS) phase-field framework \cite{kim1999phase} is utilized in this work, with a set of non-conserved order parameters (OPs) $\eta_i$ defined to distinguish between different grains in the polycrystalline Fe-doped STO  system. Within a grain interior (i.e., the bulk phase), $\eta_i = 1$ and $\eta_j = 0$ for all $j < n$ and $j \neq i$. In the GB core region between grains $i$ and $j$, the corresponding order parameters $\eta_i$ and $\eta_j$ spatially vary between 0 and 1. In this context, $\eta$ denotes the entire set of order parameters ${\eta_i}$. To properly describe the structurally and thermodynamically distinct GB core during SCL formation, we treat the GB core and the bulk as separate phases using an interpolation function $h(\eta)$ following the approach of Cha et al. \cite{cha2002phase}. When incorporating electrochemical contributions, multiple conserved concentration fields are introduced along with the electrostatic potential field to capture the distributions of charged species and the resulting electrostatic interactions. The defect concentration field variables are denoted as $C\subV$, $C\subVOne$, $C\subAccOne$, $C\subAccTwo$, $C\subAccZero$, $C\subE$, $C\subHole$ representing the concentrations of doubly ionized oxygen vacancies,  singhly ionized oxygen vacancies, singly ionzied acceptor dopants, doubly ionized acceptor dopants, neutral dopants, electrons and holes, respectively.  In addition, $\phi$ denotes the electrostatic potential. 

Then, the formulation of total free energy density functional is formulated as
\begin{equation}\label{total_density}
{F} = \int_\Omega\left[\frac{1}{2} \kappa \sum_{i=1}^n |\nabla\eta_i|^2 + \omega\sum_{i=1}^n \left(\frac{\eta_i^4}{4} - \frac{\eta_i^2}{2} + \gamma \sum_{i=1}^n \sum_{j>i}^n \eta_i^2\eta_j^2 + \frac{1}{4}\right) + f^\text{ech}\right]d\Omega,
\end{equation}
where $\kappa$ is the coefficient of the gradient term, $\omega$ is the free energy barrier coefficient. $\gamma=1.5$ ensures a symmetric profile of $\eta$ and isotropic grain growth \cite{moelans2008quantitative}. The parameters $\kappa$ and $\omega$ intrinsically determine the GB core width ($w_c$) and GB energy  ($\Gamma$) via the relations $\Gamma = \sqrt{2}/3\sqrt{\kappa\omega}$ and $w_c = \sqrt{8\kappa/\omega}$, respectively. Here, we consider only isotropic GB energy, independent of GB misorientation, because experimental evidence indicates that solute drag is the dominant factor governing AGG in Fe-doped STO (see Introduction).
$f^\text{ech}$ denotes as the electrochemical free energy density. We treat it as a mixture of bulk phase and GB core by an interpolation function $h(\eta)$. Its formulation is given by 
\begin{equation}\label{Fech}
f^\text{ech} = [1-h(\eta)]f^\text{ech}_\text{b} + h(\eta)f^\text{ech}_\text{c} - \frac{1}{2}\epsilon_\text{0}\epsilon_\text{r}(\nabla \phi)^2,
\end{equation}
where $f^\text{ech}_\text{b}$ and $f^\text{ech}_\text{c}$ are the electrochemical free energy densities of the bulk phase and GB core, respectively, and the subscripts "b" and "c" denote bulk and GB core. The last term represents the electrostatic energy density, with $\epsilon_0$ being the vacuum permittivity and $\epsilon_r$ the temperature dependent relative permittivity of the material. The interpolation function $h(\eta)$ is defined as
$h(\eta) = \frac{4}{3}\left[1-4\sum_{i=1}^n \eta_i^3 + 3\left(\sum_{i=1}^n \eta_i^2\right)^2\right]$. For a bicrystal system with a GB core located between grains $i$ and $j$, $h(\eta)$ takes the value of 0 within the grain interiors and 1 within the GB core. This interpolation function allows the bulk and GB core to be treated as distinct phases in the present model, enabling the spatially smooth assignment of different defect chemistry parameters to the bulk and GB core, in contrast to the abrupt GB$|$SCL sharp interface model, which assumes that the material constants behave as step functions between bulk and GB core \cite{de2009formation}. In the following context, we denote $h_\text{c} = h(\eta)$ and $h_\text{b} = 1 - h(\eta)$ to represent the GB core and bulk phase interpolation functions, respectively.

The electrochemical free energy densities of the bulk phase ($f^\text{ech}_\text{b}$) and GB core ($f^\text{ech}_\text{c}$) are formulated as the sum of contributions from oxygen vacancies, multivalent acceptor dopants, and electronic carriers (electrons and holes). {This additive formulation implies a non-interacting defect assumption.} The free energy densities are expressed as
\begin{equation}\label{chem_b1}
f^\text{ech}_\text{b}  =  f^\text{ech}_\text{V,b}(C\subVb, C\subVOneb) + f^\text{ech}\subAccb(C\subAccOneb, C\subAccTwob, C\subAccZerob) + f^\text{ech}\subEb(C\subEb) + f^\text{ech}\subHoleb(C\subHoleb),
\end{equation}
\begin{equation}
f^\text{ech}_\text{c}  =  f^\text{ech}_\text{V,c}(C\subVc, C\subVOnec) + f^\text{ech}\subAccc(C\subAccOnec, C\subAccTwoc, C\subAccZeroc)+ f^\text{ech}\subEc(C\subEc) + f^\text{ech}\subHolec(C\subHolec),
\end{equation}
where $C\subVb$, $C\subVc$, $C\subVOneb$, $C\subVOnec$, $C\subAccOneb$, $C\subAccOnec$, $C\subAccTwob$, $C\subAccTwoc$, $C\subAccZerob$, and $C\subAccZeroc$ denote the bulk and GB concentrations of the respective defect species. In the framework of KKS phase-field model, the total defect concentrations can be partitioned into bulk concentrations and GB concentrations. Their relationships are given by 
$C_\text{def} = h_\text{b} C_\text{def,b}  + h_\text{c} C_\text{def,c}$
with $\text{def} = \V$, $\VOne$, $\AccOne$, $\AccTwo$, $\AccZero$, $\E$ and $\Hole$. The defect concentration partitioning is based on the assumption of identical electrochemical potentials between the bulk and GB core, i.e. $\mu_\text{def} ^ \text{ech} = \mu_\text{def,b} ^ \text{ech} = \mu_\text{def,c} ^ \text{ech}$, which implies
$\mu_\text{def} ^ \text{ech} = \frac{\partial f^\text{ech}_\text{b}}{\partial C_\text{def,b}} = \frac{\partial f^\text{ech}_\text{c}}{\partial C_\text{def,c}}.$

Each individual electrochemical free energy contribution ($f^\text{ech}_\text{V,b}$, $f^\text{ech}_\text{V,c}$, $f^\text{ech}\subAccb$, $f^\text{ech}\subAccc$, $f^\text{ech}\subEb$, $f^\text{ech}\subEc$, $f^\text{ech}\subHoleb$, and $f^\text{ech}\subHolec$) is formulated to remain consistent with defect chemistry theory. In this framework, multivalent acceptor dopants ($\AccOne$, $\AccTwo$ and $\AccZero$) occupy B-sites of the perovskite sublattice, substituting for the host B-site cations (e.g. Ti in STO), while singly and doubly oxygen vacancies ($\VOne$ and $\V$) correspond to missing oxygen atoms on the O-site. Importantly, B-site acceptor dopants and O-site oxygen vacancies belong to different sub-lattices and thus must be accounted for separately, in order to appropriately represent the respective configurational entropies. Moreover, the local structure within the GB core differs from that of the bulk. These structural differences can significantly alter the number of available sites for both dopants and oxygen vacancies. Hence, the numbers of available sites of oxygen vacancies multivalent acceptor dopants in bulk phase and GB core, denoted as ${N}_\text{V,b}$,  ${N}_\text{V,c}$, ${N}\subAccb$, ${N}\subAccc$ should be explicitly distinguished in the chemical energy contributions. Additionally, the concentrations of electrons and holes are intrinsically limited by the effective densities of states ($N_\text{CB}$ and $N_\text{VB}$), which define the maximum carrier concentrations in the CBM and VBM,, respectively.
Therefore, the individual electrochemical energy densities in dilute solution limit are defined as follows
\begin{equation}\label{Fech_Vb}
\begin{split}
f^\text{ech}_\text{V,b} &= \mu^0_\text{V,b} (C\subVb + C\subVOneb)+ \bc T\left[ C\subVb\ln\left(\frac{C\subVb}{{N}_\text{V,b}}\right)+ C\subVOneb\ln\left(\frac{C\subVOneb}{{N}_\text{V,b}}\right) \right.\\
&\left. + ({N}_\text{V,b}-C\subVb-C\subVOneb)\ln\left(\frac{{N}_\text{V,b}-C\subVb - C\subVOneb}{{N}_\text{V,b}}\right)\right] + z\subV e C\subVb \phi + z\subVOne e C\subVOneb \phi,
\end{split}
\end{equation}
\begin{equation}\label{Fech_Vc}
\begin{split}
f^\text{ech}_\text{V,c} &= \mu^0_\text{V,c} (C\subVc + C\subVOnec)+ \bc T\left[ C\subVc\ln\left(\frac{C\subVc}{{N}_\text{V,c}}\right)+ C\subVOnec\ln\left(\frac{C\subVOnec}{{N}_\text{V,c}}\right) \right.\\
&\left. + ({N}_\text{V,c}-C\subVc-C\subVOnec)\ln\left(\frac{{N}_\text{V,c}-C\subVc - C\subVOnec}{{N}_\text{V,c}}\right)\right] + z\subV e C\subVc \phi + z\subVOne e C\subVOnec \phi,
\end{split}
\end{equation}
\begin{equation}
\begin{split}
f^\text{ech}\subAccb = &\mu^0\subAccb(C\subAccOneb + C\subAccTwob + C\subAccZerob) + \bc T \left[C\subAccOneb\ln\left(\frac{C\subAccOneb}{{N}\subAccb}\right) + C\subAccTwob\ln\left(\frac{C\subAccTwob}{{N}\subAccb}\right) + C\subAccZerob\ln\left(\frac{C\subAccZerob}{{N}\subAccb}\right) \right.\\
&\left. + ({N}\subAccb-C\subAccOneb-C\subAccTwob-C\subAccZerob)\ln\left(\frac{{N}\subAccb-C\subAccOneb-C\subAccTwob-C\subAccZerob}{{N}\subAccb}\right) \right] \\
&+ z\subAccOne e C\subAccOneb \phi + z\subAccTwo e C\subAccTwob \phi +  z\subAccZero e C\subAccZerob \phi ,
\end{split}
\end{equation}
\begin{equation}
\begin{split}
f^\text{ech}\subAccc = &\mu^0\subAccc(C\subAccOnec + C\subAccTwoc + C\subAccZeroc) + \bc T \left[C\subAccOnec\ln\left(\frac{C\subAccOnec}{{N}\subAccc}\right) + C\subAccTwoc\ln\left(\frac{C\subAccTwoc}{{N}\subAccc}\right) + C\subAccZeroc\ln\left(\frac{C\subAccZeroc}{{N}\subAccc}\right) \right.\\
&\left. + ({N}\subAccc-C\subAccOnec-C\subAccTwoc-C\subAccZeroc)\ln\left(\frac{{N}\subAccc-C\subAccOnec-C\subAccTwoc-C\subAccZeroc}{{N}\subAccc}\right) \right]\\
&+ z\subAccOne e C\subAccOnec \phi + z\subAccTwo e C\subAccTwoc \phi +  z\subAccZero e C\subAccZeroc \phi ,
\end{split}
\end{equation}
\begin{equation}
f^\text{ech}\subEb= \mu^0\subEb C\subEb + \bc T\left[ C\subEb\ln\left(\frac{C\subEb}{N_\text{CB,b}}\right) + (N_\text{CB,b}-C\subEb)\ln\left(\frac{N_\text{CB,b}-C\subEb}{N_\text{CB,b}}\right)\right] + z\subE e C\subEb \phi,
\end{equation}
\begin{equation}
f^\text{ech}\subEc= \mu^0\subEc C\subEc + \bc T\left[ C\subEc\ln\left(\frac{C\subEc}{N_\text{CB,c}}\right) + (N_\text{CB,c}-C\subEc)\ln\left(\frac{N_\text{CB,c}-C\subEc}{N_\text{CB,c}}\right)\right] + z\subE e C\subEc \phi,
\end{equation}
\begin{equation}
f^\text{ech}\subHoleb= \mu^0\subHoleb C\subHoleb + \bc T\left[ C\subHoleb\ln\left(\frac{C\subHoleb}{N_\text{VB,b}}\right) + (N_\text{VB,b}-C\subHoleb)\ln\left(\frac{N_\text{VB,b}-C\subHoleb}{N_\text{VB,b}}\right)\right] + z\subHole e C\subHoleb \phi,
\end{equation}
\begin{equation}\label{Fech_Hc}
f^\text{ech}\subHolec= \mu^0\subHolec C\subHolec + \bc T\left[ C\subHolec\ln\left(\frac{C\subHolec}{N_\text{VB,c}}\right) + (N_\text{VB,c}-C\subHolec)\ln\left(\frac{N_\text{VB,c}-C\subHolec}{N_\text{VB,c}}\right)\right] + z\subHole e C\subHolec \phi,
\end{equation}
Here, $\mu^0_\text{V,b}$, $\mu^0_\text{V,c}$, $\mu^0\subAccb$, $\mu^0\subAccc$, $\mu^0\subEb$, $\mu^0\subEc$, $\mu^0\subHoleb$, and $\mu^0\subHolec$ are the standard formation energies of oxygen vacancies, acceptor dopants, electrons, and holes in the bulk and GB core. The difference of the standard formation energy of oxygen vacancies ($\Delta\mu_\text{V} = \mu^0_\text{V,c} - \mu^0_\text{V,b}$) triggers the SCL formation, while other formation energy difference ($\Delta\mu\subAcc = \mu^0\subAccc - \mu^0\subAccb$, $\Delta\mu\subE = \mu^0\subEc - \mu^0\subEb$, $\Delta\mu\subHole = \mu^0\subHolec - \mu^0\subHoleb$) can change the SCL profiles. $\bc$ is the Boltzmann constant, $T$ is the temperature, and $e$ is the elementary charge. $z\subV$, $z\subVOne$, $z\subAccOne$, $z\subAccTwo$, $z\subAccZero$, $z\subE$, and $z\subHole$ are the valence states of doubly ionized oxygen vacancies, singly ionized oxygen vacancies, singly ionized acceptor dopants, doubly ionized acceptor dopants, neutral dopants, electrons, and holes, respectively.  Then,  the electrochemical potentials of different charged species in bulk and GB core are derived from the electrochemical free energy (see Appendix).

\subsection{Evolution equations} \label{evolution_equations}
 
In the present work, the GB anisotropy is not included. The GB energy and mobility are constant. Then, the phase-field OPs for grain $i$ with isotropic GB properties are evolved using Allen-Cahn equations
\be \label{AC_equation}
\frac{1}{L}\frac{\partial \eta}{\partial t} = - \frac{\delta F}{\delta \eta},
\ee
where $L$ is the constant GB mobility. Substituting \Cref{total_density} into \Cref{AC_equation}, we have
\be \label{PF_equation1}
    \frac{1}{L}\frac{\partial \eta_i}{\partial t} = \nabla\cdot\kappa\nabla\eta_i - \omega \frac{\partial f^{\text{loc}}(\eta)}{\partial \eta_i} + \frac{\partial h(\eta)}{\partial \eta_i}\left[ f_\text{b}^{\text{ech}} - f_{\text{c}}^{\text{ech}} - \sum_{\text{def}} \frac{\partial f_\text{b}^{\text{ech}}}{\partial C_\text{def,b}} (C_\text{def,b} - C_\text{def,c})\right],
\ee
with $\text{def} = \VOne, \V, \AccOne, \AccTwo, \AccZero, \E, \Hole$. For a moving GB core with a constant velocity ($v_\text{GB}$) along $x$ direction, we have $\frac{\partial \eta_i}{\partial t} = v_\text{GB} \frac{\partial \eta_i}{\partial x}$, then \Cref{PF_equation1} becomes
\be
\frac{v_\text{GB}}{L}\frac{\partial \eta_i}{\partial x} = \nabla\cdot\kappa\nabla\eta_i - \omega \frac{\partial f^{\text{loc}}(\eta)}{\partial \eta_i} + \frac{\partial h(\eta)}{\partial \eta_i}\left[ f_\text{b}^{\text{ech}} - f_{\text{c}}^{\text{ech}} - \sum_{\text{def}} \frac{\partial f_\text{b}^{\text{ech}}}{\partial C_\text{def,b}} (C_\text{def,b} - C_\text{def,c})\right].
\ee

The concentrations of charged oxygen vacancies and multivalent acceptor dopants are treated as independent field variables and evolve according to coupled reaction–diffusion equations \cite{chen2022classical, lei2021phase}. The diffusive flux of each species is driven by its electrochemical potential gradient, while local charge-state transitions are described by mass-action-law reaction terms. For the singly and doubly ionized oxygen vacancies ($\VOne$ and $\V$), the corresponding concentration evolution equations are given by
\be \label{V_Con_eq}
\frac{\partial C\subV}{\partial t} = \nabla  \left[M\subV\left(\nabla\mu\subV^\text{ech}\right)\right] + \nu\subV,
\ee
\be \label{VOne_Con_eq}
\frac{\partial C\subVOne}{\partial t} = \nabla  \left[M\subVOne\left(\nabla\mu\subVOne^\text{ech}\right)\right] + \nu\subVOne,
\ee
 where $M\subVOne$ and $M\subV$ are the mobilities of singly and doubly ionized oxygen vacancies, which are given by $M\subV = D_\text{V}/(\partial^2 f^\text{ech}/\partial C\subV^2)$ and $M\subVOne = D_\text{V}/(\partial^2 f^\text{ech}/\partial C\subVOne^2)$,  with $D_\text{V}$ being the diffusivities of oxygen vacancy. {The diffusivities of doubly charged oxygen vacancies and of singly charged oxygen vacancies in STO are expected to be different, since in the first case the oxide ion (O$^{2-}$) jumps into an empty vacancy whereas in the second case it jumps into a vacancy that holds an electron. Data for for oxygen vacancies in HfO$_2$ \cite{duncan2016filament, mueller2021importance} confirm this picture, but for STO, no such data are available. We assume, therefore, for simplicity that the two defects have identical mobilities.}
$\nu\subVOne$ and $\nu\subV$ denote as the net reaction rates of singly and doubly ionized oxygen vacancies through defect reaction \eqref{Rred1}. The formation of $\nu\subVOne$ and $\nu\subV$ are
\be
\nu\subVOne = -k_{\text{V}_\text{O}1\rightarrow 2}^\text{f} C\subVOne + k_{\text{V}_\text{O}1\rightarrow 2}^\text{b} C\subV C\subE,
\ee
\be
\nu\subV = k_{\text{V}_\text{O}1\rightarrow 2}^\text{f} C\subVOne - k_{\text{V}_\text{O}1\rightarrow 2}^\text{b} C\subV C\subE,
\ee
where $k_{\text{V}_\text{O}1\rightarrow 2}^\text{f}$ and $k_{\text{V}_\text{O}1\rightarrow 2}^\text{b}$ denote the forward and backward rate constants of the defect reaction \eqref{Rred1}. At thermodynamically equilibrium state, we have $\nu\subVOne = \nu\subV = 0$,  leading to the equilibrium relation 
\be
K_{\text{V}_\text{O}1\rightarrow 2} = \frac{k_{\text{V}_\text{O}1\rightarrow 2}^\text{f}}{k_{\text{V}_\text{O}1\rightarrow 2}^\text{b}}.
\ee

For multivalent acceptor dopants, in addition to diffusion driven by electrochemical potential gradients, charge-state transitions are described by the ionization reactions in \Cref{Rion1} and \Cref{Rion2}. These reactions interconvert the neutral, singly ionized, and doubly ionized dopant states ($\AccZero$, $\AccOne$, and $\AccTwo$) through the capture or release of electron holes. As a result, the local concentration distributions of multivalent acceptor dopants are governed by the coupled effects of diffusion and ionization kinetics.
Accordingly, the evolution equations for the multivalent acceptor dopants can be written as
\be \label{AccOne_Con_eq}
\frac{\partial C\subAccOne}{\partial t} = \nabla  \left[M\subAccOne\left(\nabla\mu\subAccOne^\text{ech}\right)\right] + \nu\subAccOne,
\ee
\be \label{AccTwo_Con_eq}
\frac{\partial C\subAccTwo}{\partial t} = \nabla  \left[M\subAccTwo\left(\nabla\mu\subAccTwo^\text{ech}\right)\right] + \nu\subAccTwo,
\ee
\be \label{AccZero_Con_eq}
\frac{\partial C\subAccZero}{\partial t} = \nabla  \left[M\subAccZero\left(\nabla\mu\subAccZero^\text{ech}\right)\right] + \nu\subAccZero,
\ee
with $M\subAccOne$, $M\subAccTwo$, $M\subAccZero$ being the mobilities of the multivalent acceptor dopants, given by $M\subAccOne = D\subAcc/(\partial^2 f^\text{ech}/\partial C\subAccOne^2)$, $M\subAccTwo = D\subAcc/(\partial^2 f^\text{ech}/\partial C\subAccTwo^2)$ and $M\subAccZero = D\subAcc/(\partial^2 f^\text{ech}/\partial C\subAccZero^2)$. For simplicity, all dopant charge states are also assumed to share the same diffusivity $D\subAcc$. $\nu\subAccOne$, $\nu\subAccTwo$ and $\nu\subAccZero$ denote the net reaction rates associated with the ionization reactions \Cref{Rion1} and \Cref{Rion2}, expressed as the difference between forward and backward reaction terms:
\be
\nu\subAccZero = - k_\text{ion1}^\text{f} C\subAccZero +  k_\text{ion1}^\text{b} C\subAccOne C\subHole,
\ee
\be
\nu\subAccTwo = k_\text{ion2}^\text{f} C\subAccOne - k_\text{ion2}^\text{b} C\subAccTwo C\subHole,
\ee
\be
\nu\subAccOne = (k_\text{ion1}^\text{f} C\subAccZero - k_\text{ion1}^\text{b} C\subAccOne C\subHole) + (-k_\text{ion2}^\text{f} C\subAccOne  + k_\text{ion2}^\text{b} C\subAccTwo C\subHole). 
\ee
The forward and backward reaction rate constants $k_\text{ion1}^\text{f}$, $k_\text{ion1}^\text{b}$, $k_\text{ion2}^\text{f}$, and $k_\text{ion2}^\text{b}$ are constrained by thermodynamic consistency. {In the present work, these reaction rates are assumed to be sufficiently large compared to the characteristic diffusion rates, such that charge-state transitions occur on a much faster timescale than defect diffusion.} At equilibrium, where $\nu\subAccZero = \nu\subAccOne = \nu\subAccTwo = 0$, the following relationships hold:
\begin{equation}
K_\text{ion1} = \frac{k_\text{ion1}^\text{f}}{k_\text{ion1}^\text{b}}, \quad
K_\text{ion2} = \frac{k_\text{ion2}^\text{f}}{k_\text{ion2}^\text{b}}.
\end{equation}
The equilibrium defect reaction constants $K_\text{ion1}$ and $K_\text{ion2}$ depend on temperature and the CTLs. Specifically, they are given by: 
\begin{equation}
K_\text{ion1} = N_\text{VB} \exp\left(-\frac{E_\text{A1}}{k_B T}\right), \quad
K_\text{ion2} = N_\text{VB} \exp\left(-\frac{E_\text{A2}}{k_B T}\right),
\end{equation}
{where $E_{\text{A1}}$ and $E_{\text{A2}}$ denote the energy separations between the $\mathrm{Fe}^{4+/3+}$ and $\mathrm{Fe}^{3+/2+}$ CTLs and the VBM, respectively.}
{As temperature increases, the band gap decreases due to its intrinsic temperature dependence. This may lead to an unphysical situation where $E_{\text{A2}}$ exceeds the band gap, i.e., the $\mathrm{Fe}^{3+/2+}$ CTL would lie above the CBM.
To avoid this inconsistency, we assume that the energy separation between the CBM and $\mathrm{Fe}^{3+/2+}$ CTL remains constant with temperature. Specifically, this separation is taken to be 0.27~eV at 0~K and is maintained at all temperatures considered in this work for Fe-doped STO. Hence, $E_{\text{A2}}$ in the present work is temperature dependent, and its expression is $E_{\text{A2}}(T) = E_\text{g}(T)-0.27$. Here, $E_{\text{g}}(T) = E_{\text{CB}} - E_{\text{VB}}$ is the temperature-dependent band gap. $E_{\text{VB}}$ and $E_{\text{CB}}$ denote the energies of the VBM and CBM, respectively.}

Then, CTLs enter the acceptor dopant concentration evolution equations not merely as equilibrium descriptors, but as fundamental parameters that shape the local reaction kinetics. At equilibrium state, the vanishing of the acceptor dopant evolution terms requires the elimination of both electrochemical potential gradients and net ionization reaction rates, i.e., $\nu\subAccZero = \nu\subAccOne = \nu\subAccTwo = 0$. 
In non-equilibrium conditions, such as during SCL formation or GB migration, dopant charge-state transitions become highly sensitive to the local alignment between the Fermi level and the CTLs. These transitions introduce temporally evolving source or sink terms in the dopant evolution equations, thereby locally enhancing or suppressing the rate of dopant redistribution. 

Additionally, for electronic carriers (electrons and holes), assumptions can be made to simplify the model by reducing the number of independent parameters. For instance, the electronic structure, and thus the effective density of states and standard formation energies of electrons and holes, is not expected to vary significantly between the bulk and GB core regions. Therefore, it is reasonable to assume that $\mu^0\subEb = \mu^0\subEc$, $\mu^0\subHoleb = \mu^0\subHolec$, $N_\text{CB,b} = N_\text{CB,c} = N_\text{CB}$, and $N_\text{VB,b} = N_\text{VB,c} = N_\text{VB}$. Under these assumptions, the electrochemical potentials of electrons and holes in the bulk and GB core become equal when their concentrations satisfy $C\subEb = C\subEc = C\subE$ and $C\subHoleb = C\subHolec = C\subHole$. In addition, due to the extremely high mobilities of electrons and holes, their electrochemical potentials can be considered to be constant ($E_\text{F}$) within the SCLs. Therefore, the concentrations of electrons and holes can be expressed as:
\begin{equation} \label{Ce_EF}
C\subE = N_\text{CB}\exp\left(\frac{E_\text{F} - E_\text{CB} - z\subE e\phi}{\bc T}\right),
\end{equation}
\begin{equation}\label{Ch_EF}
C\subHole = N_\text{VB}\exp\left(\frac{E_\text{VB} - E_\text{F} - z\subHole e\phi}{\bc T}\right),
\end{equation}
where $E_\text{F}$ is the Fermi energy. The product of of electron concentration and hole concentration is fixed to be $C\subE C\subHole = N_\text{CB}N_\text{VB}\exp\left(-\frac{E_\text{g}}{\bc T}\right)$ \cite{sze2021physics}.

Finally, the electrostatic potential governing equation is given by the variational derivative of the total free energy density functional with respect to $\phi$ \cite{garcia2004thermodynamically}:
\be
\frac{\delta F}{\delta \phi} = 0.
\ee
Then we obtain
\be
\epsilon_0\epsilon_r\nabla^2\phi + \rho = 0,
\ee
where $\rho$ is the charge density, and is given by
\be
\begin{split}
\rho =& z\subVOne C\subVOne + z\subV C\subV + z\subAccOne C\subAccOne + z\subAccTwo C\subAccTwo + z\subE C\subE + z\subHole C\subHole \\
=& z\subVOne \left(C\subVOneb h_\text{b} + C\subVOnec h_\text{c}\right) + z\subV \left(C\subVb h_\text{b} + C\subVc h_\text{c}\right) + z\subAccOne \left(C\subAccOneb h_\text{b} + C\subAccOnec h_\text{c} \right)\\
&+ z\subAccTwo \left(C\subAccTwob h_\text{b} + C\subAccTwoc h_\text{c} \right) + z\subE C\subE + z\subHole C\subHole.
\end{split}
\ee

\subsection{Bulk defect concentrations determination by defect chemistry and charge transition levels}
CTLs not only influence the acceptor dopant redistribution dynamics, but also determine the bulk defect concentrations at initial stage.
To determine the equilibrium bulk concentrations of these defects far from the GB core ($x = \infty$), two fundamental constraints are applied. First, the constraint of electroneutrality is given by
\be \label{electroneutrality_constrain}
\begin{split}
&z\subVOne C\subVOneb(\infty) + z\subV C\subVb(\infty) + z\subAccOne C\subAccOneb(\infty) + z\subAccTwo C\subAccTwob(\infty) + z\subAccZero C\subAccZerob(\infty)\\
& + z\subE C\subEb(\infty) + z\subHole C\subHoleb(\infty) = 0,
\end{split}
\ee
and second, the mass conservation law for the acceptor dopants,
\be \label{mass_conservation}
C\subAccOneb(\infty) +  C\subAccTwob(\infty) + C\subAccZerob(\infty) = C\subAccb(\infty),
\ee
where $C\subAccb(\infty)$ denotes the bulk doping level, corresponding to the total concentration of the acceptor dopants in the bulk. 
In addition to these constraints, the defect concentrations follow the defect reactions \eqref{Rred}–\eqref{Rion2} and are related through their respective mass action laws:
\be \label{Kred}
C\subVb(\infty)\pO^{1/2}C^2\subE(\infty) = K_\text{red},
\ee
\be \label{Kred1}
\frac{C\subV(\infty) C\subE(\infty)}{C\subVOne(\infty)} = K_{\text{V}_\text{O}1\rightarrow 2},
\ee
\be \label{Kion1}
\frac{C\subAccOneb(\infty)C\subHole(\infty)}{C\subAccZerob(\infty)} = K_\text{ion1},
\ee
\be \label{Kion2}
\frac{C\subAccTwob(\infty)C\subHole(\infty)}{C\subAccOneb(\infty)} = K_\text{ion2},
\ee
\be \label{Keh}
C\subE(\infty)C\subHoleb(\infty) =  K_\text{eh},
\ee
where $K_\text{red}$, $K_{\text{V}_\text{O}1\rightarrow 2}$, $K_\text{eh}$, $K_\text{ion1}$, and $K_\text{ion2}$ are the equilibrium constants of the defect reactions \eqref{Rred}, \eqref{Rred1}, \eqref{Reh}, \eqref{Rion1} and \eqref{Rion2}, respectively, which depend on temperature and the oxygen partial pressure. 
For specific temperatures and oxygen partial pressures, the bulk concentrations $C\subVOneb(\infty)$, $C\subVb(\infty)$, $C\subAccOneb(\infty)$, $C\subAccTwob(\infty)$, $C\subAccZerob(\infty)$, $C\subE(\infty)$, and $C\subHole(\infty)$ can be obtained by solving Eqs. \eqref{electroneutrality_constrain} to \eqref{Keh}. These bulk defect concentrations are used to initialize the following phase-field simulations.  After obtaining the defect concentration in the bulk, the Fermi energy levels can be determined. We assume the equilibrium states of electrons and holes are always satisfied due to their high mobilities. Hence, the electrochemical potentials of electrons and holes are constant in the simulation domain, i.e. $\mu^\text{ech}\subE = \mu^{0}\subE + \bc T\ln(C\subE(\infty)/N_\text{CB}) + z\subE e \phi(\infty)= E_\text{F}$. With $\phi(\infty)=0$, we have
\be
E_\text{F} = E_\text{g} + \bc T\ln\left[\frac{C\subE(\infty)}{N_\text{CB}}\right].
\ee  
The Fermi level is constant everywhere in the simulations and determine the concentrations of electrons and holes via \eqref{Ce_EF} and \eqref{Ch_EF}.

\begin{figure}[h]
    \centering
    \includegraphics[width=0.9\linewidth]{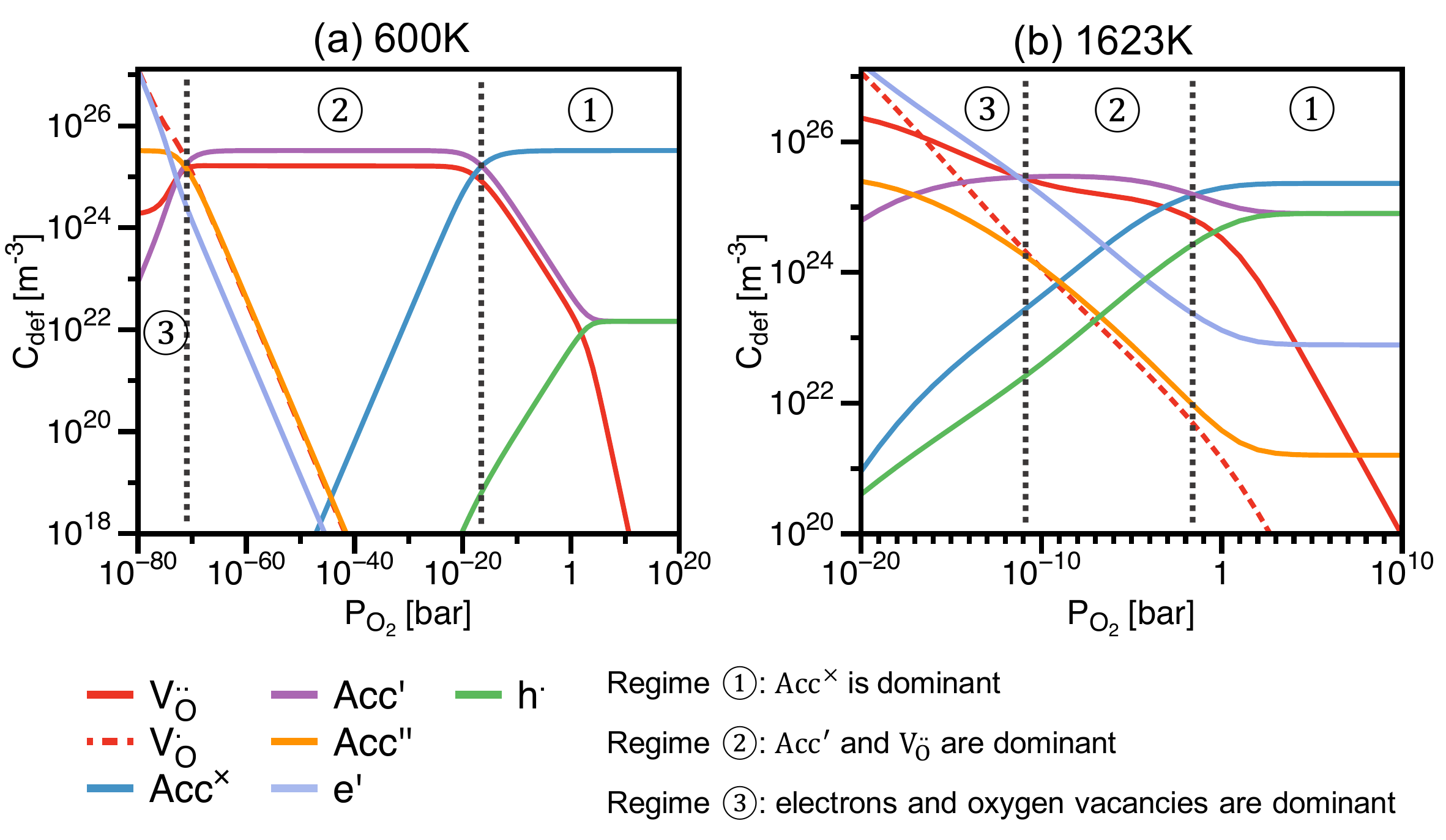}
    \caption{ Calculated bulk defect concentrations for 0.2\% Fe-doped STO at 600 K and 1623 K under varying oxygen partial pressure. Three distinct regimes can be identified based on the dominant defects: {Regime 1} — neutral acceptor dopants dominate at high oxygen partial pressures; {Regime 2} — singly ionized acceptor dopants and oxygen vacancies dominate at intermediate oxygen partial pressures; {Regime 3} — oxygen vacancies and electrons dominate at low oxygen partial pressures.}
    \label{figure1}
\end{figure}

The degrees-of-freedoms in the present phase-field model are summarized in \Cref{DOF}. For Fe-doped STO, the defect chemistry parameters are taken from Moos et al. \cite{moos1997defect} and listed in \Cref{ParameterTable}. 
The calculated bulk defect concentrations for 0.2\% Fe-doped STO are shown in \Cref{figure1} at two representative temperatures (600~K and 1623~K) under various oxygen partial pressures.
Based on the dominant defect species across different oxygen partial pressures, three characteristic regimes can be identified:
\begin{itemize}
  \item Regime 1: At high oxygen partial pressures (oxidizing conditions), neutral acceptor dopants are the predominant defect species, and the Fermi level is located between the VBM and the Fe$^{4+/3+}$ CTL. In this regime, the fraction of singly ionized dopants relative to neutral ones in the bulk, far from the GB core, converges to a finite value (e.g., $\sim$0.0015 at 600~K and $\sim$0.35 at 1623~K), indicating that singly ionized acceptor dopants cannot completely convert into neutral ones under oxidizing conditions, particularly at the sintering temperature. Therefore, considering the influence of neutral acceptor dopants on SCL formation in this regime is of crucial importance. In addition, singly ionized acceptor dopants are compensated either by oxygen vacancies or by holes depending on the temperature, and the system exhibits $p$-type conduction behavior.
  
\item Regime 2: At intermediate oxygen partial pressures, the concentrations of singly ionized acceptor dopants and oxygen vacancies increase significantly due to progressive ionization of neutral dopants. The Fermi level shifts upward, crossing the Fe$^{3+/4+}$ CTL. This regime is therefore characterized by stronger electrostatic interactions and enhanced ionic compensation compared to regime~1.

\item Regime 3: At low oxygen partial pressures (reducing condition), the dominant defects shift to oxygen vacancies and electrons. The Fermi level moves closer to the CBM, indicating a transition to $n$-type semiconducting behavior driven by electron accumulation. Meanwhile, the concentration of neutral acceptor dopants becomes significantly lower than that of their ionized counterparts, including both singly and doubly ionized states.

\end{itemize}

\section{Numerical implementation of the CUDA-based phase-field framework}

In this section, we describe the defect chemistry parameters selection and the numerical implementation of the phase-field model, which includes the governing equations for the OPs, defect concentrations, and electrostatic potential. 
The physical parameters used to simulate the Fe-doped STO are listed in  \Cref{ParameterTable}. The formation energy difference of oxygen vacancy between GB core and bulk is -1.5~eV, which is obtained via atomistic simulations. The details of the atomistic simulations are presented in Supplementary. 

The governing equations for the phase-field order parameters and conserved defect concentrations are discretized using finite difference methods in space and explicit time integration schemes \cite{leveque2007finite}. 
The simulations are carried out using a Compute Unified Device Architecture (CUDA) based parallel computing framework to achieve high computational efficiency on graphics processing unit (GPU) \cite{sanders2010cuda}.
One-dimensional computational domain is discretized into a regular grid with a uniform grid spacing of $\Delta x=0.2 w_\text{c} $, where $w_\text{c}$ denotes the GB core width. The numbers of grid points in the x-directions are set to 10,240, to simulate a bicrystal configuration with the GB core located at the center of the simulation domain.
All field variables, including the non-conserved order parameters, conserved defect concentration fields, and electrostatic potential, are defined and initialized at the grid nodes. The initialization of defect concentrations under different temperatures and oxygen partial pressures follows the bulk defect chemistry as shown in \Cref{figure1}.
 To ensure numerical stability and computational efficiency, an adaptive time-stepping algorithm is employed based on the total free energy reduction rate \cite{ZHANG2013204adaptive}. The electrostatic potential is calculated by solving the Poisson equation, which is discretized using finite differences and solved iteratively via the Gauss-Seidel method \cite{leveque2007finite}.
The CUDA implementation involves designing parallel kernels to execute the following tasks efficiently:
\begin{itemize}
\item Calculation of spatial derivatives, such as the Laplacian terms (e.g., $\nabla^2 \eta$), using five-point finite difference stencils and evaluation of local free energy densities and chemical potentials at each grid point.
\item Time stepping updates for the phase-field order parameters ($\eta$) and defect concentration fields ($C\subVOne$, $C\subV$, $C\subAccOne$, $C\subAccTwo$, and $C\subAccZero$).
\item Calculation of electron and hole concentrations ($C\subE$ and $C\subHole$), charge density, and solution of the Poisson equation using Gauss-Seidel iterative solver to obtain the electrostatic potential ($\phi$) with the relative tolerance for convergence being $1\times 10^{-10}$.
\item Analytical calculation of the phase concentrations in the bulk and GB core based on the KKS phase-field model, by enforcing the equality of chemical potentials between the bulk and GB core to determine the partitioned concentrations ($C\subVOneb$, $C\subVOnec$, $C\subVb$, $C\subVc$, $C\subAccOneb$, $C\subAccOnec$, $C\subAccTwob$, $C\subAccTwoc$, $C\subAccZerob$, $C\subAccZeroc$).
\end{itemize}

Boundary conditions are imposed either by explicitly treating the boundary nodes within the CUDA kernels, depending on the requirements of the specific governing equation. For the phase-field order parameters, Dirichlet boundary conditions are applied. For defect concentrations, Neumann boundary conditions are used to ensure zero flux at the boundaries. For the electrostatic potential, the left side of the simulation domain is grounded (Dirichlet condition), while the right side applies a Neumann condition to maintain a zero normal derivative.

In this work, the CUDA-based phase-field framework is implemented in C++ and executed on NVIDIA Tesla A100 GPUs in the Lichtenberg high-performance computing cluster at Technische Universit\"at Darmstadt. Each simulation takes approximately 20 hours to reach the equilibrium state.
This computational cost arises from the high complexity of the problem. Compared to previous defect-chemistry-informed phase-field models, which only consider doubly ionized oxygen vacancies and singly ionized acceptor dopants \cite{wang2024defect}, the present model includes multiple defect species with different charge states as well as the electrostatic potential field. As a result, the number of degrees of freedom is significantly increased, leading to a larger set of coupled equations to solve.
Consequently, numerous iterations are required to reach the thermodynamic equilibrium state due to the strong coupling between the phase-field variables, defect concentrations, and electrostatics. Although finite difference methods are efficient and straightforward for time-dependent phase-field simulations, they become less efficient when only the equilibrium state is of interest, because of the small time step constraints and the slow convergence near the energy minimum in the later stages of the simulations. Nevertheless, with the aid of an adaptive time-stepping scheme and the high computational efficiency enabled by the CUDA-based GPU implementation, the simulations can be carried out within reasonable computational times while maintaining numerical stability and accuracy. Furthermore, this CUDA-based phase-field framework can be readily extended to simulate three-dimensional microstructure evolution during grain growth in polycrystalline systems.

\begin{table}[H]
    \centering
    \caption{Summarized degree of freedoms in the present phase-field model}
    \label{DOF}
    \begin{tabular}{llc}
\hline   \textbf{Degrees-of-freedom} & \textbf{Definition} & \textbf{Unit} \\
\hline      $\eta_i$   &   phase-field order parameter of grain $i$, $i=1,2,3,...,n$   &  -      \\  
$C\subVOne$   &  Local concentration of singly ionized oxygen vacancies   &  $\text{m}^{-3}$     \\  
$C\subVOneb$   &  Concentration of singly ionized oxygen vacancies in bulk phase   &  $\text{m}^{-3}$     \\ 
$C\subVOnec$   &  Concentration of singly ionized oxygen vacancies in GB core   &  $\text{m}^{-3}$     \\ 
$C\subV$   &  Local concentration of doubly ionized oxygen vacancies   &  $\text{m}^{-3}$     \\  
$C\subVb$   &  Concentration of doubly ionized oxygen vacancies in bulk phase   &  $\text{m}^{-3}$     \\ 
$C\subVc$   &  Concentration of doubly ionized oxygen vacancies in GB core   &  $\text{m}^{-3}$     \\ 
$C\subAccZero$   &  Local concentration of neutral acceptor dopants  &  $\text{m}^{-3}$     \\  
$C\subAccZerob$   & Concentration of neutral acceptor dopants in bulk phase  &  $\text{m}^{-3}$     \\  
$C\subAccZeroc$   &  Concentration of neutral acceptor dopants in GB core &  $\text{m}^{-3}$     \\  
$C\subAccOne$   &  Local concentration of singly ionized acceptor dopants  &  $\text{m}^{-3}$     \\
$C\subAccOneb$   &  Concentration of singly ionized acceptor dopants in bulk phase  &  $\text{m}^{-3}$     \\
$C\subAccOnec$   &  Concentration of singly ionized acceptor dopants in GB core  &  $\text{m}^{-3}$     \\
$C\subAccTwo$   &  Local concentration of doubly ionized acceptor dopants  &  $\text{m}^{-3}$     \\  
$C\subAccTwob$   &  Concentration of doubly ionized acceptor dopants in bulk phase &  $\text{m}^{-3}$     \\  
$C\subAccTwoc$   & Concentration of doubly ionized acceptor dopants in GB core &  $\text{m}^{-3}$     \\  
$C\subE$   &  Local concentration of electrons  &  $\text{m}^{-3}$     \\  
$C\subHole$   &  Local concentration of holes  &  $\text{m}^{-3}$     \\  
${\phi}$   & Electrostatic potential & V \\
\hline   
\end{tabular}  
\end{table}

\begin{table}[]
    \centering
    \caption{Summarized physical and defect chemistry parameters used in the present work for Fe-doped STO.}
    \label{ParameterTable}
    \begin{tabular}{llll}
        \hline
        {Parameters} & {Value} & Unit & {Source} \\
        \hline
        $T$ & 600 and 1623 & K & - \\
        $a$ & $3.9\times 10^{-10}+6.64\times 10^{-15}T/K$& $\text{m}$ & \cite{kemp2024one} \\
        $w_\text{c}$ & 2a & $\text{m}$ & - \\
        $N_\text{V,b}$  & $3/a^3 $ &$\text{m}^{-3}$ & \cite{usler2024space} \\
        $N\subAccb$  & $1/a^3 $ &$\text{m}^{-3}$ & \cite{usler2024space}  \\
        $N_\text{V,c}$  & $1\times 10^{27}$  &$\text{m}^{-3}$ & \cite{de2009formation}\\
        $N\subAccc$  & $1\times 10^{27}$ &$\text{m}^{-3}$ & \cite{de2009formation}  \\
        $C\subAccb(\infty)$ & 0.002$N\subAccb$ &$\text{m}^{-3}$ & - \\
         $\Delta\mu_\text{V}$  & -1.5 & eV & atomistic simulations (SI)  \\
          $\Delta\mu\subAcc$  & -1.5 & eV &  \cite{de2009formation} \\
           $\Delta\mu\subE$  & 0 & eV &  - \\
            $\Delta\mu\subHole$  & 0 & eV &  - \\
        $K_\text{red}$  & $5\times 10^{89}\exp\left(-\frac{6.1~\text{eV}}{\bc T}\right)$ &$\text{m}^{-9}\text{bar}^{1/2}$ & \cite{moos1997defect} \\
           $K_{\text{V}_\text{O}1\rightarrow 2}$  & $N_\text{CB}\exp\left(-\frac{E_{\text{V}_\text{O}1\rightarrow 2}}{\bc T}\right)$ &$\text{m}^{-3}$ & \cite{moos1997defect} \\
        $K_\text{eh}$  & $N_\text{CB}N_\text{VB}\exp\left(-\frac{E_\text{g}}{\bc T}\right)$ &$\text{m}^{-6}$ &  \cite{moos1997defect} \\
        $K_\text{ion1}$ & $N_\text{VB}\exp\left(-\frac{E_\text{A1}}{\bc T}\right)$ &$\text{m}^{-3}$ & \cite{moos1997defect} \\
        $K_\text{ion2}$ & $N_\text{CB}\exp\left(-\frac{E_\text{A2}}{\bc T}\right)$ &$\text{m}^{-3}$ &  \cite{moos1997defect, maier2016low2} \\
        $E_\text{g}$ & $3.17 -5.66\times10^{-4} (T/\text{K}) $ &eV &  \cite{moos1997defect} \\
        $N_\text{CB}$ & $4.1\times 10^{22}(T/\text{K})^{1.5}$ &$\text{m}^{-3}$ &  \cite{moos1997defect} \\
         $N_\text{VB}$ & $3.5\times 10^{22}(T/\text{K})^{1.5}$ &$\text{m}^{-3}$ &  \cite{moos1997defect} \\
         $E_{\text{V}_\text{O}1\rightarrow 2}$ & 0.3 &eV   &  \cite{moos1997defect} \\
         $E_\text{A1}$ & 0.94 &eV   &  \cite{moos1997defect} \\
         $E_\text{A2}$ & $E_\text{g}-0.27$ &eV   &  \cite{suzuki2019energy} \\
         $u\subV$ & $1.0\times 10^8 T^{-1} \exp\left(\frac{-0.86}{\bc T}\right)$ &  $\text{m}^{2}\text{V}^{-1}\text{s}^{-1}$  &  \cite{denk1995partial, meyer2003observation} \\
         $u\subE$ & $3.95\times 10^8 T^{-1.62} $ &  $\text{m}^{2}\text{V}^{-1}\text{s}^{-1}$  &  \cite{moos1997defect} \\
         $u\subHole$ & $1.1\times 10^{10} T^{-2.36} $ &  $\text{m}^{2}\text{V}^{-1}\text{s}^{-1}$  &  \cite{moos1997defect} \\
         $z\subVOne$ & +1 &  -  &  - \\
          $z\subV$ & +2 &  -  &  - \\
          $z\subAccOne$ & -1 &  -  &  - \\
          $z\subAccTwo$ & -2 &  -  &  - \\
          $z\subAccZero$ & 0 &  -  &  - \\
          $z\subE$ & -1 &  -  &  - \\
          $z\subHole$ & +1 &  -  &  - \\
         $D_\text{V}$ & $\mu\subV \bc T/ z\subV e$  & $\text{m}^{2}/\text{s}$ &  -\\
         $D\subAcc$ & $0.01 D_\text{V}$ & $\text{m}^{2}/\text{s}$   & \cite{vikrant2020electrochemical} \\
         $\epsilon_{\text{0}}$ & $8.85\times 10^{-12}$ & $\text{C/(Vm)}$   & - \\
         $\epsilon_{\text{r}}$ & $90000/(T-35)$ & -   & \cite{de2009formation} \\
         $\bc$ & $8.617\times 10^{-5}$ & $\text{eV K}^{-1}$   & - \\
           $\Gamma$ & $ 0.6 $ & $ \text{J}\text{m}^{-2}$   & \cite{vikrant2020electrochemical} \\
        \hline
    \end{tabular}
\end{table}

\section{Results and discussions}

\subsection{Charge transition level effects on space charge layer formation at stationary grain boundaries}

In this section, we examine the equilibrium SCL formation in 0.2\% Fe-doped STO under different thermodynamic conditions at stationary GBs. Phase-field simulations are carried out at two representative temperatures, 600~K and 1623~K.  For each temperature, a wide range of oxygen partial pressures is explored to investigate how CTLs and defect equilibria influence the spatial distribution of charged defects and the electrostatic potential profile across the GB region in both materials.

At 600~K, the acceptor dopants are assumed to be frozen-in, and the redistribution of their charge states is governed primarily by hole transport during ionization reactions. The SCL formation is simulated across oxygen partial pressures ranging from $10^{-10}$ to $10^{-80}$~bar. For Fe-doped STO, we focus on three representative cases with $P_{\mathrm{O}_2} = 10^{-15}$, $10^{-60}$, and $10^{-75}$~bar, corresponding to regimes 1, 2, and 3, respectively. 

At 1623~K, the acceptor dopants are assumed to be mobile and able to reach full thermodynamic equilibrium, where their spatial redistribution is controlled by both diffusion and ionization reaction processes. In this high-temperature regime, the SCL formation is simulated over an oxygen partial pressure range from $10^{-14}$ to $10^{6}$~bar. In this case, we focus on three representative cases at $P_{\mathrm{O}_2} = 10^{2}$, $10^{-6}$, and $10^{-14}$~bar, which correspond to regimes 1, 2, and 3, respectively. 

The phase-field simulation results at 600~K and 1623~K are presented in \Cref{STO_EQSCL}. The spatial distributions of charged defect concentrations, electrostatic potential, and the corresponding electronic band structure across the SCL are illustrated. In the band structure plots, the energy level of the VBM in the bulk region (far from the GB core) is taken as the reference ($E_{\mathrm{VB}}(\infty) = 0$). Within the SCL, the VBM edge shifts with the local electrostatic potential according to $E_{\mathrm{VB}}(x) = -\phi(x)$, while the conduction band edge follows $E_{\mathrm{CB}}(x) = E_{\mathrm{g}} - \phi(x)$, where $E_{\mathrm{g}}$ is the band gap and $\phi(x)$ is the local electrostatic potential obtained from the phase-field simulations.

\subsubsection{Frozen-in acceptor dopant at 600~K}

In \Cref{STO_EQSCL}(a), the phase-field results of equilibrium SCL formation at 600~K are demonstrated. In regime 1, exemplified by $P_{\mathrm{O}_2} = 10^{-15}$~bar and shown in \Cref{STO_EQSCL}, neutral acceptor dopants 
dominate in Fe-doped STO. The singly ionized acceptor dopants 
are compensated by doubly ionized oxygen vacancies. Although the total dopant distribution is assumed to be frozen-in and initially uniform, the simulation reveals the formation of a local accumulation zone of singly ionized acceptor dopants near the GB. In the GB core, the concentration of singly ionized acceptor dopants become significantly higher than that of the neutral acceptor dopants without diffusion process, indicating a charge-state transition from Fe$^{4+}$ to Fe$^{3+}$ has occurred. This transition is corroborated by the electronic band structure. As shown in the band diagram, the Fermi level (approximately 0.91~eV) remains spatially constant across the SCL. In the bulk region, it lies slightly below the Fe$^{4+/3+}$ CTL (0.94~eV), while in the SCL, the Fe$^{4+/3+}$ CTL bends downward due to the local electrostatic potential and falls below the Fermi level. This alignment facilitates the reduction of Fe$^{4+}$ to Fe$^{3+}$ within the GB region. We also calculate the areal charge density in the GB core as $Q_\text{c} = \int^{w_\text{c}/2}_{-w_\text{c}/2} (e\sum_\text{def} z_\text{def}c_\text{def}) \text{d}x$. At $P_{\mathrm{O}_2} = 10^{-15}$~bar, the area charge density in the GB core is 0.118~$\text{C}/\text{m}^2$.

In regime 2, corresponding to $P_{\mathrm{O}_2} = 10^{-60}$~bar, the dominant defect species in Fe-doped STO remain singly ionized acceptor dopants and doubly ionized oxygen vacancies. Compared to regime 1, the spatial distribution of singly ionized acceptor dopants become nearly uniform across the SCL. Under these conditions, the Fermi level lies at approximately 2.2~eV, approaching the CBM (2.83~eV), which leads to a noticeable increase in electron concentration across the SCL.
In addition, this regime exhibits distinct redox behavior. The simulation results show that the concentration of doubly ionized acceptor dopants becomes significant within the GB core. This is attributed to the Fermi level intersecting with the Fe$^{3+/2+}$ CTL, thereby enabling further reduction from Fe$^{3+}$ to Fe$^{2+}$. Doubly ionized acceptor dopants, being more negatively charged, offers a stronger compensating effect for the positively charged oxygen vacancies in the GB core. The area charge density in the GB core is 0.137~$\text{C}/\text{m}^2$.

In regime 3, corresponding to an extremely low oxygen partial pressure (e.g., $P_{\mathrm{O}_2} = 10^{-75}$~bar), singly ionized oxygen vacancies and electrons are the dominant defect species, and Fe-doped STO exhibits pronounced n-type semiconducting behavior. Under such strongly reducing conditions, the concentration of oxygen vacancies increases dramatically, and the Fermi level shifts upward to approximately 2.7~eV, approaching the CBM. The dominant charge state of the acceptor dopant is $\text{Fe}_\text{Ti}''$, which has a nearly uniform distribution across SCL. In this case, the area charge density in the GB core is 0.0724~$\text{C}/\text{m}^2$.

\subsubsection{Mobile acceptor dopant at 1623~K}
\begin{figure}[]
    \centering
    \includegraphics[width=0.9\linewidth]{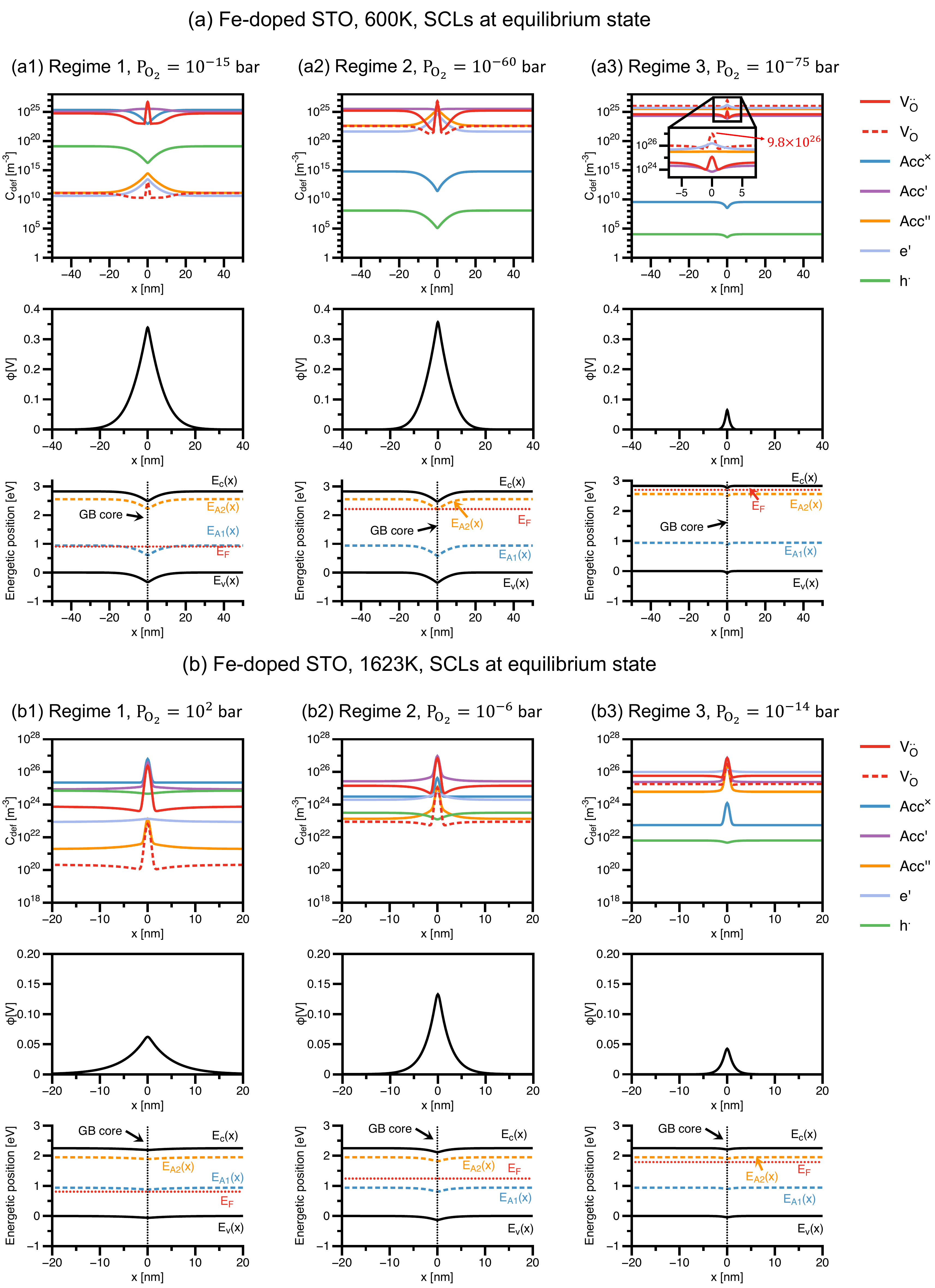}
    \caption{The equilibrium SCL formation in bicrystalline 0.2\% Fe-doped STO at (a) 600 K and (b) 1623 K under various oxygen partial pressures is shown. For each temperature, the defect-concentration profiles, electrostatic-potential distribution, and band diagrams across SCL in regimes 1, 2, and 3 are presented.}
    \label{STO_EQSCL}
\end{figure}

In contrast to the frozen-in acceptor dopants at 600~K, both diffusion and ionization processes govern the concentration distribution of acceptor dopants at 1623~K. The phase-field results are plotted in \Cref{STO_EQSCL}(b). At this elevated temperature, band gap narrowing causes the CBM to shift downward, thereby modifying the relative alignment between the Fermi level and the CTLs. 


At $10^2$~bar (regime~1), the Fermi level (0.81~eV) lies slightly below the Fe$^{4+/3+}$ CTL (0.94~eV), such that neutral acceptor dopants constitute the predominant defect species. The fraction of singly ionized dopants relative to neutral ones in the bulk is approximately 0.38, indicating that only a minor portion of dopants can effectively contribute to SCL formation. Charge compensation in this regime is therefore provided mainly by this limited population of singly ionized acceptor dopants in combination with electron holes, which results in a comparatively low GB potential and weak CTL bending. Under this condition, the area charge density in the GB core is 0.0111~$\text{C}/\text{m}^2$. 

At $10^{-6}$~bar (regime~2), the Fermi level shifts upward to 1.24~eV, and approximately 90\% of the acceptor dopants are singly ionized. As a consequence, the SCL is dominated by singly ionized acceptor dopants, with doubly ionized oxygen vacancies serving as the principal compensating species. In this regime, the GB potential increases significantly and the CTL bending becomes more pronounced. Accumulation of singly and doubly ionized dopants is observed in the SCL adjacent to the GB core, whereas neutral dopants exhibit negligible spatial variation in the SCL. The calculated area charge density in the GB core is 0.0421~$\text{C}/\text{m}^2$. 

At $10^{-14}$~bar (regime~3), the Fermi level (1.78~eV) approaches the CBM, and doubly ionized oxygen vacancies together with electrons become the dominant defects in the bulk. Within the SCL, however, the positively charged oxygen vacancies is compensated not only by electrons but also by singly ionized acceptor dopants. As a result, the GB potential in this regime is reduced compared with that at $10^{-6}$~bar. Under this condition, the area charge density in the GB core is 0.0236~$\text{C}/\text{m}^2$. 

Additionally, singly and doubly ionized acceptor dopants exhibit accumulation zones within the SCL near the GB core in all three regimes, whereas the distribution of neutral dopants remains nearly uniform in the vicinity of GB core, which is distinct from the pronounced depletion zone observed at 600~K when acceptor dopants are frozen-in. The absence of a depletion zone for neutral acceptor dopants at high temperatures can be attributed to two factors. First, neutral dopants are uncharged and therefore do not interact directly with the electrostatic potential of the SCL. Second, although the GB potential promotes charge-state transitions between neutral and singly ionized dopants, this process is strongly coupled to diffusion at high temperature. Consequently, the redistribution of neutral acceptor dopants is governed by the combined effects of ionization reactions and diffusion, which lead to a homogenized spatial profile and suppress the formation of a depletion zone.

\subsubsection{Charge transition level effects on the space charge layer profile}

The influence of CTL on the SCL profile at stationary GBs arises from two coupled effects. First, the position of the CTLs with respect to the Fermi level determines the equilibrium charge states of acceptor dopants in the bulk. Second, within the SCL, where the electrostatic potential is spatially inhomogeneous, CTLs govern the local ionization reactions and thereby control the redistribution of dopant charge states. As a result, the relative alignment of the Fermi level and CTLs dictates not only the bulk defect concentrations but also the spatial variation of charged species and the electrostatic potential across the GB. 

\begin{figure}[h]
    \centering
    \includegraphics[width=0.9\linewidth]{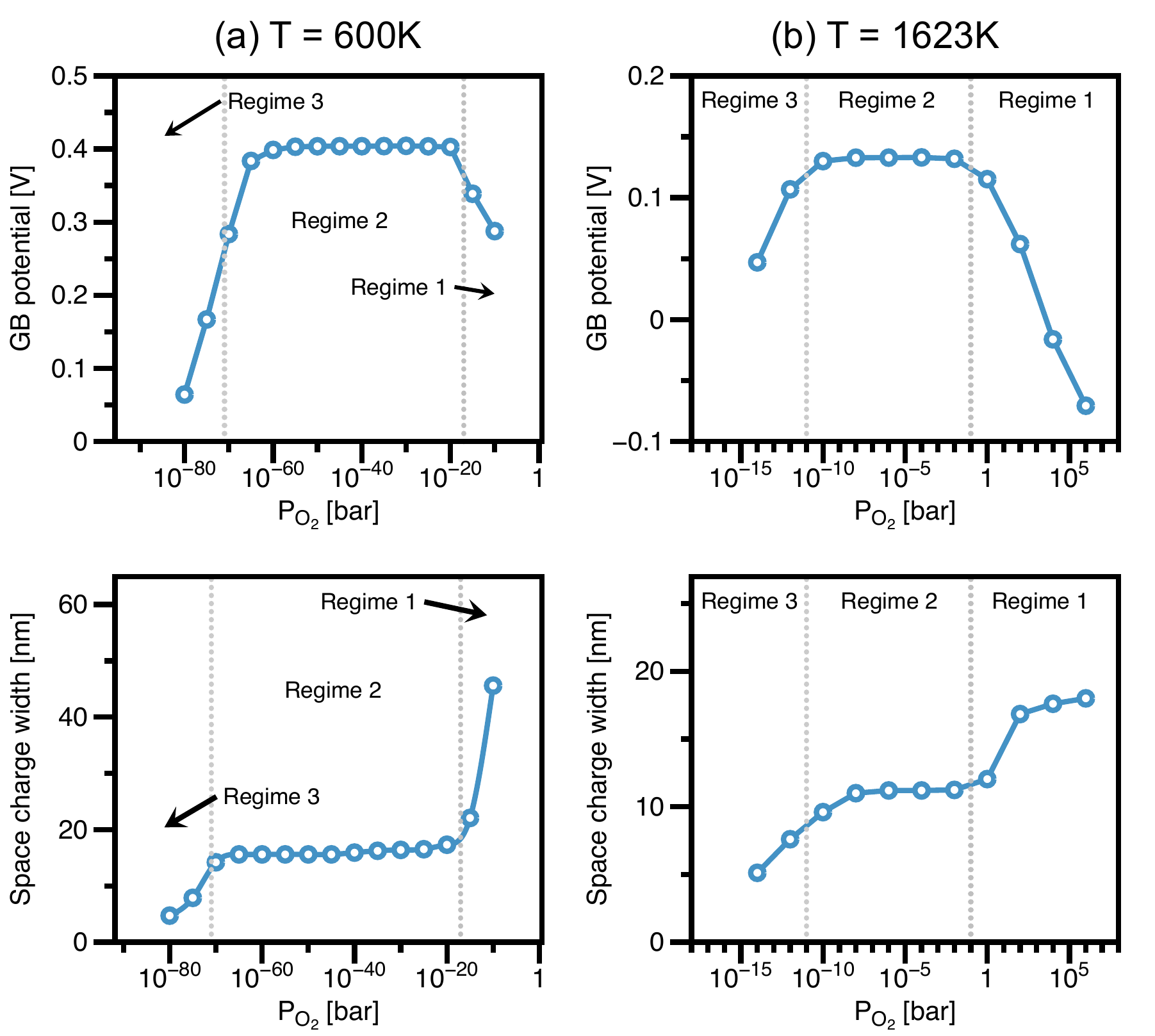}
    \caption{The GB potential and the space-charge width in Fe-doped STO at (a) 600~K and (b) 1623~K as a function of oxygen partial pressure. The GB potential is extracted at the GB core ($\eta = 0.5$) in the phase-field simulations. The space-charge width is extracted from the chemical width and estimated by visual inspection of the oxygen vacancies (600~K) and acceptor dopants (1623~K) concentration profiles \cite{usler2024space}.}
    \label{Figure_space_charge_potential}
\end{figure}

In \Cref{Figure_space_charge_potential}, we demonstrate the effects of CTLs on the GB potential and the space-charge width as functions of oxygen partial pressure at 600~K and 1623~K. At 600~K, the GB potential reaches its maximum in regime~2, whereas it decreases significantly in regimes~1, 3, and 4. 
In regime~2, the concentration of oxygen vacancies becomes comparable to that of singly ionized acceptor dopants. However, when the acceptor dopants are frozen-in, the oxygen vacancies segregated at the GB core cannot be effectively compensated by negatively charged species, leading to an enhanced electrostatic potential across the SCL. 
When the oxygen partial pressure increases to regime~1, neutral acceptor dopants become the dominant species. In this regime, the concentrations of both oxygen vacancies and singly ionized acceptor dopants are substantially reduced. As a result, the number of oxygen vacancies segregated at the GB core decreases, and the oxygen vacancies can be partially compensated by the increased singly ionized dopants generated through Fe$^{4+/3+}$ charge-state transitions induced by CTL bending within the SCL. 
In regime~3, although the concentrations of oxygen vacancies and electrons are much higher than those of singly ionized acceptor dopants, the segregated oxygen vacancies at the GB core can be largely compensated by fast-moving electrons. This strong electronic screening markedly reduces the GB potential compared with regime~2.
At 1623~K, the GB potential also peaks in regime~2 where the Fermi level is above the Fe$^{4+/3+}$ CTL, maximizing the fraction of singly ionized acceptor dopants, and decreases toward both more oxidizing (regime~1, dominated by neutral dopants) and more reducing conditions (regime~3, where oxygen vacancies and electrons dominate). The overall similarity reflects that the governing mechanism, the alignment of the Fermi level with the Fe$^{4+/3+}$ CTL, remains unchanged across temperatures.

Not only the GB potential but also the space-charge width is influenced by charge transition effects. At high temperature, the space-charge width can be described within the Gouy–Chapman model as \cite{gregori2017ion}
\be
\lambda_\text{GC} = \sqrt{\frac{\epsilon_0\epsilon_r \bc T}{2z_\text{maj}^2 e^2 C_\text{maj,b}(\infty)}},
\ee
where $C_\text{maj,b}(\infty)$ is the concentration of the majority mobile defects in the bulk and $z_\text{maj}$ is their charge number. At lower temperatures, when dopant mobility is suppressed, the Mott–Schottky model applies, yielding
\be
\lambda_\text{MS} = \lambda_\text{GC}\sqrt{\frac{4e\Phi_\text{0}}{\bc T}},
\ee
with $\Phi_\text{0}$ denoting the GB potential \cite{gregori2017ion}. Clearly, a larger $C_{\text{maj},b}(\infty)$ leads to a shorter space-charge width.  The relative alignment of the Fermi level with the CTLs directly governs the value of $C_{\text{maj},b}(\infty)$ and thereby controls the space-charge width. In regime~1, the much smaller bulk concentration of singly ionized acceptor dopants leads to longer space charge layer at 600~K and 1623~K. In regime~2, the increase in singly ionized acceptor dopants substantially raises $C_{\text{maj},b}(\infty)$, thereby reducing the space-charge width compared with regime~1.
In regime~3, the space-charge width reaches its minimum. Under strongly reducing conditions, the Fermi level approaches the conduction-band minimum, leading to a pronounced increase in the concentrations of oxygen vacancies and electrons, which far exceed those of the acceptor dopants. The extremely high densities of these charged species result in the narrowest space-charge width among all regimes.

\subsection{Charge transition level effects on space charge layer formation and solute drag behavior during grain boundary migration}

While CTL-induced charge-state transitions of acceptor dopants govern the equilibrium SCL profiles at stationary GBs, their role during GB migration remains barely understood. During sintering, GB migration gives rise to solute drag effects due to the low diffusivity of acceptor dopants, resulting in asymmetric SCL formation \cite{wang2024defect} and asymmetric CTL bending. Such asymmetry shifts the relative alignment of the Fermi level with the CTLs, thereby modifying local charge-state transitions within the SCL and, in turn, altering the strength of the solute drag effect.

In this section, we investigate how CTLs influence SCL formation during GB migration, with particular emphasis on their coupling to solute drag under different GB migration velocities ($v_\text{GB}$) at typical sintering temperatures (e.g., 1623~K). In \Cref{QE_SCL1}, \ref{QE_SCL2} and \ref{QE_SCL3}, we examine the impact of the Fe\textsuperscript{4+/3+} CTL on asymmetric SCL formation under GB velocities spanning slow to fast migration. Further details on the influence of Fe\textsuperscript{4+/3+} CTL bending on solute drag behavior are provided in \Cref{solute_drag_force}. 

\subsubsection{Charge transition level effects on asymmetric space charge layer formation}

\begin{figure}[]
    \centering \includegraphics[width=1\linewidth]{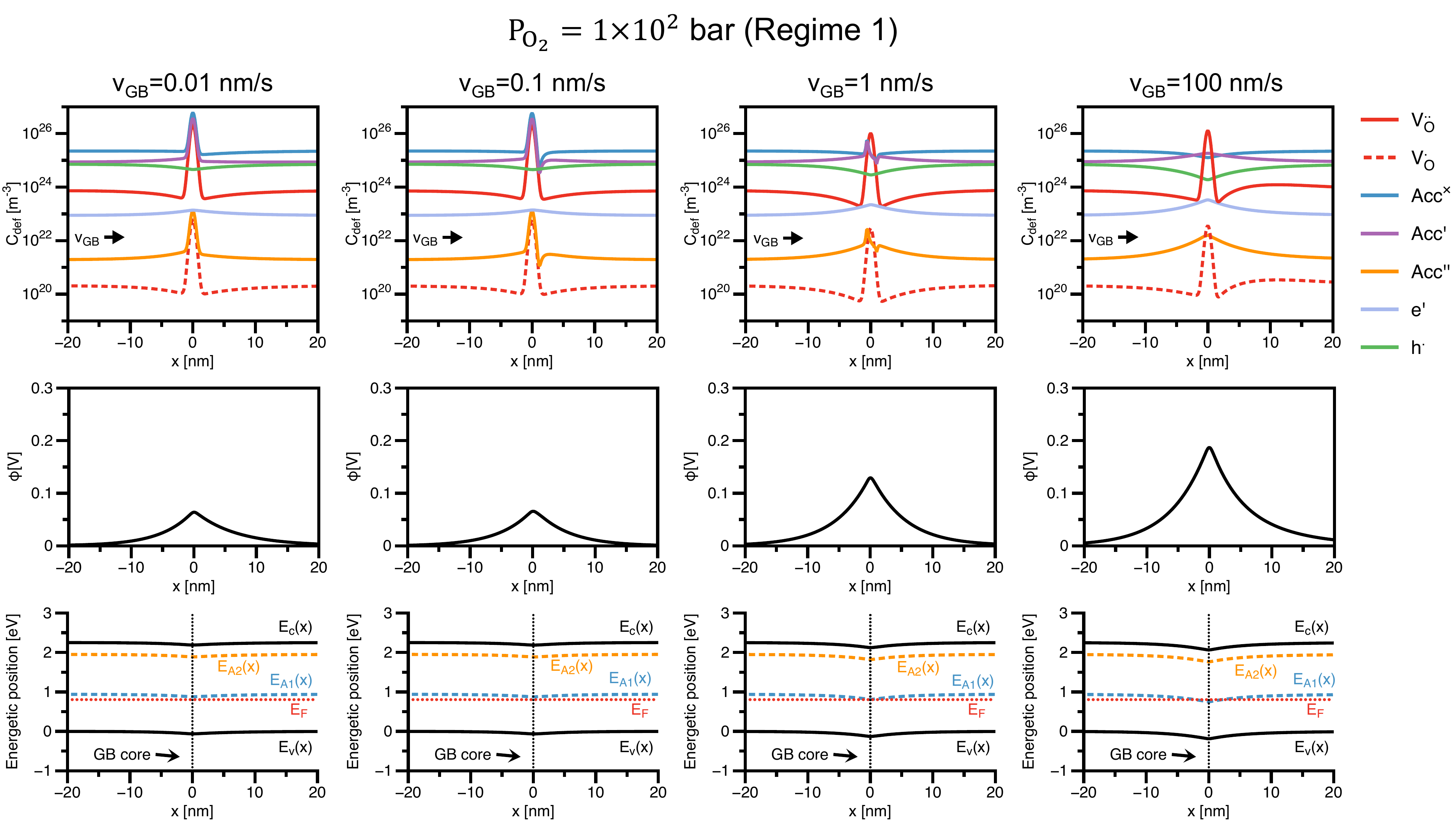}
    \caption{Asymmetric SCL formation at quasi-equilibrium state for different GB moving velocities for $P_{\text{O}_2} = 1\times 10^{2}$~bar (Regime~1).}
    \label{QE_SCL1}
\end{figure}

\begin{figure}[]
    \centering \includegraphics[width=1\linewidth]{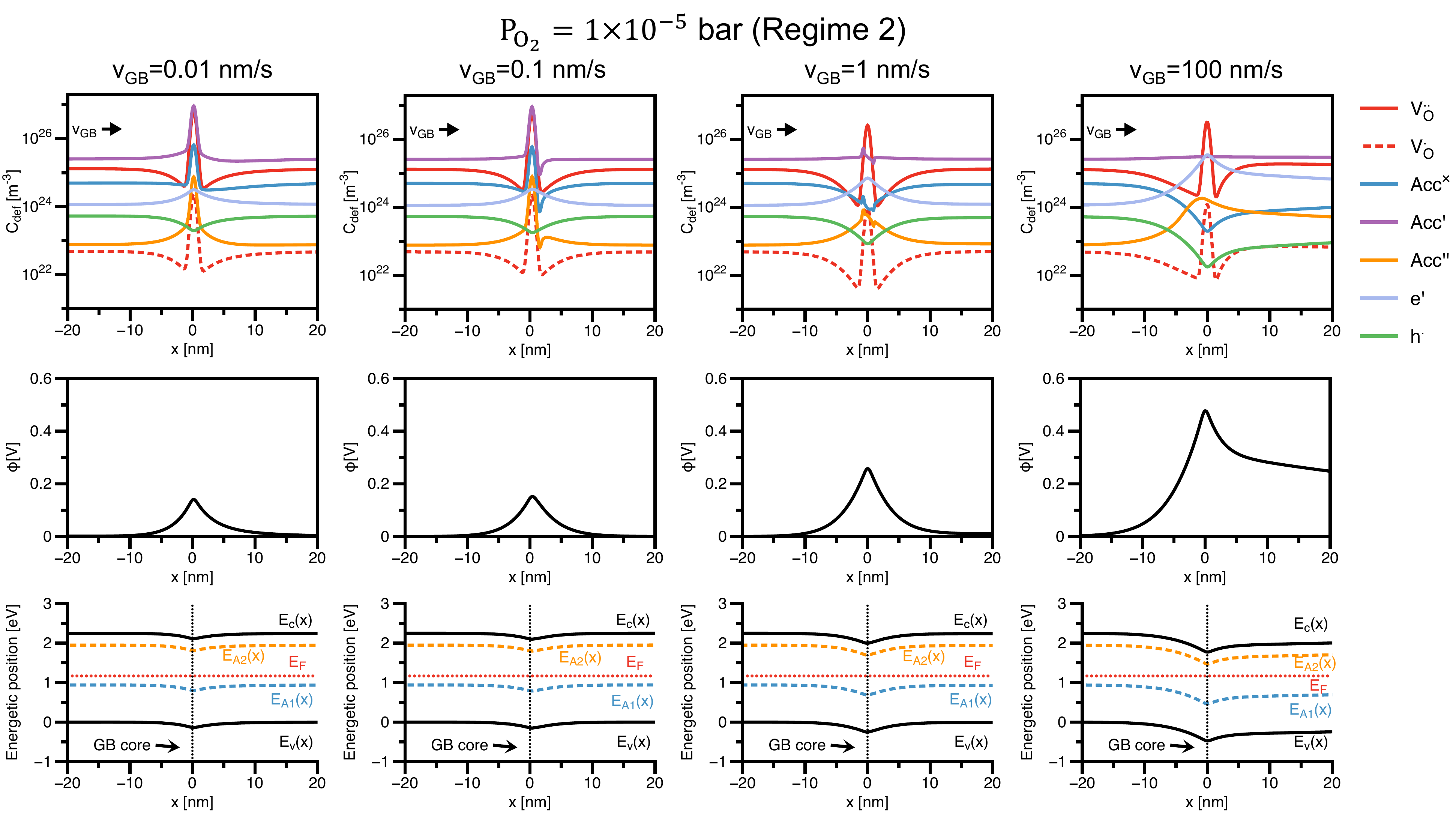}
    \caption{Asymmetric SCL formation at quasi-equilibrium state for different GB moving velocities for $P_{\text{O}_2} = 1\times 10^{-5}$~bar (Regime~2).}
    \label{QE_SCL2}
\end{figure}

\begin{figure}[]
    \centering \includegraphics[width=1\linewidth]{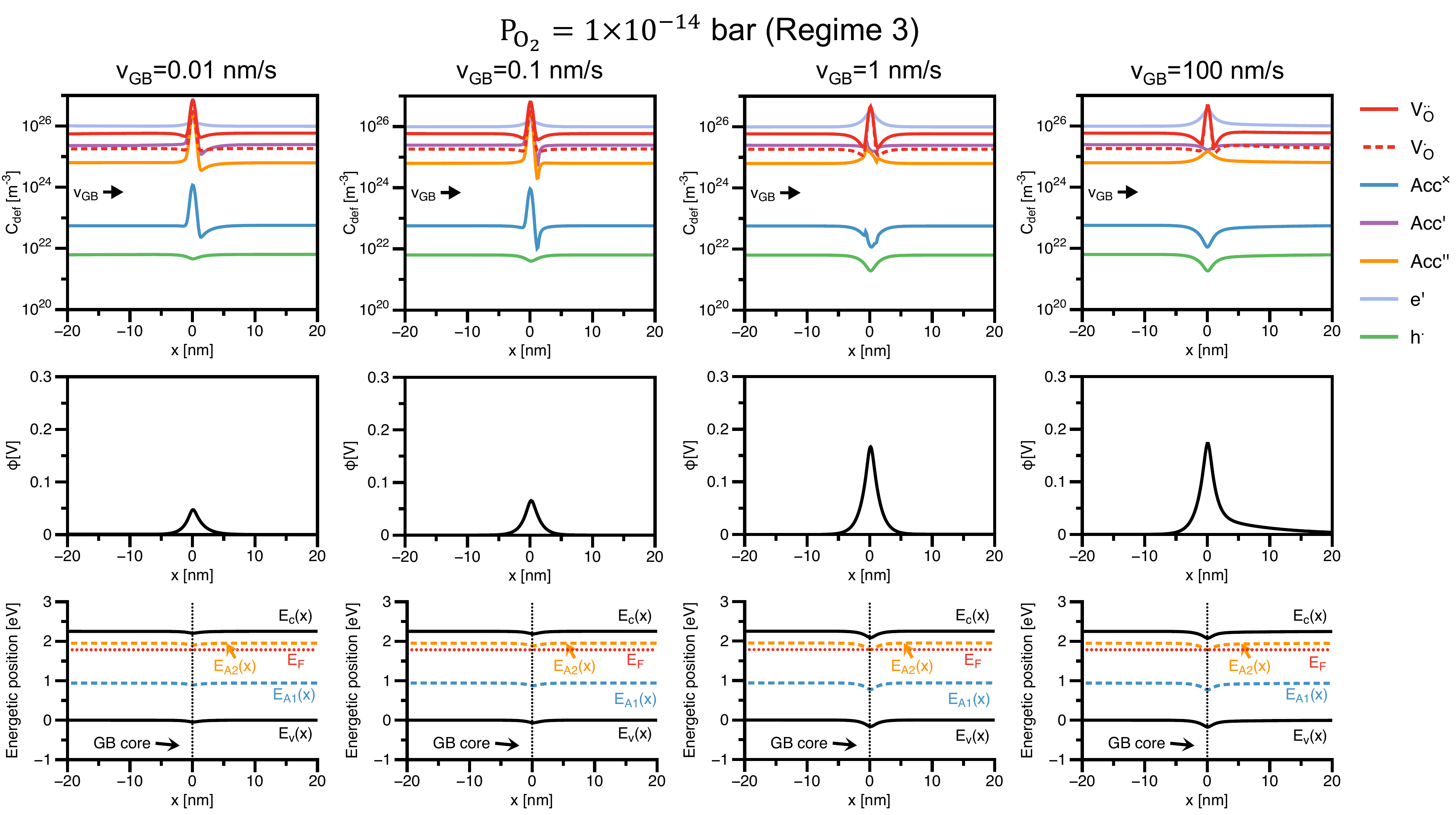}
    \caption{Asymmetric SCL formation at quasi-equilibrium state for different GB moving velocities for $P_{\text{O}_2} = 1\times 10^{-14}$~bar (Regime~3).}
    \label{QE_SCL3}
\end{figure}

The asymmetric SCL formation, as shown in \Cref{QE_SCL1}, \ref{QE_SCL2} and \ref{QE_SCL3} is evaluated under different oxygen partial pressures of $10^2$, $10^{-5}$, and $10^{-14}$~bar, combined with GB velocities of 0.01, 0.1, 1, and 100~nm/s.
In regime~1 (see \Cref{QE_SCL1}), represented by $P_{\text{O}_2} = 10^2$~bar, the Fermi level is below the Fe$^{4+/3+}$ CTL, and neutral acceptor dopants are the dominant species. At a slow GB velocity of 0.01~nm/s, asymmetric SCL formation begins to appear. As the GB velocity increases to 0.1 or 1~nm/s, the asymmetry becomes more pronounced, and the accumulation zone of acceptor dopants can even transform into a depletion zone ahead of the propagating GB core (see the region at $x > 0$~nm).
At very high GB velocities, such as 100~nm/s, the low diffusivity of acceptor dopants prevents them from following the GB migration, and segregation at the GB core is strongly suppressed. With fewer dopants available for charge compensation, positively charged oxygen vacancies are less compensated, resulting in a higher electrostatic potential within the SCL. This velocity-dependent increase in GB potential also enhances Fe$^{4+/3+}$ CTL bending. At 0.01~nm/s, the Fermi level does not intersect the Fe$^{4+/3+}$ CTL, whereas at 100~nm/s, the Fe$^{4+/3+}$ CTL lies below the Fermi level, particularly in the GB core region, thereby promoting charge-state transitions from neutral to singly ionized acceptor dopants.

In regime~2, the Fermi level lies above the Fe$^{4+/3+}$ CTL, and singly ionized acceptor dopants dominate the defect chemistry. With increasing GB velocity, asymmetric SCL formation is also observed. At high GB velocities, the distribution of singly ionized acceptor dopants becomes nearly uniform.
Compared with regime~1, however, the electrostatic potential distribution exhibits pronounced asymmetry. The space-charge width ahead of the propagating GB is significantly larger (exceeding 100~nm), giving rise to a strongly asymmetric SCL profile. This asymmetry enhances the bending of the Fe$^{4+/3+}$ CTL, which promotes distinct charge-state transitions from neutral to singly ionized acceptor dopants across the GB, and consequently leads to a marked reduction in the concentration of neutral acceptor dopants in front of the moving GB core.

In regime~3, oxygen vacancies and electrons dominate the defect chemistry, and the Fermi level lies close to the conduction-band minimum. At low GB velocities, the distribution of acceptor dopants becomes asymmetric. However, owing to the high diffusivity of oxygen vacancies and electrons, their spatial asymmetry across the SCL remains limited. At high GB velocities, the electrostatic potential increases and the SCL on the forward side of the migrating boundary becomes significantly thicker. This not only alters the distribution of electrons (see Eq.\eqref{Ce_EF}), but also enhances CTL bending, which in turn modifies the charge-state transitions of acceptor dopants.

Therefore, CTLs play a critical role in modulating asymmetric SCL formation during GB migration. Because of their very low diffusivity, acceptor dopants cannot keep pace with the migrating boundary, resulting in an asymmetric electrostatic potential across the SCL. In all three regimes, higher GB velocities lead to increased GB potentials, which in turn cause stronger bending of the Fe$^{4+/3+}$ CTL and promote the conversion of neutral dopants into singly ionized species. In addition, increasing GB velocity produces a thicker SCL region on the forward side of the migrating boundary (particularly in regime~2, where the space-charge width exceeds 100~nm), thereby extending the spatial range of CTL bending and further enhancing charge-state transitions from neutral to singly ionized acceptors.
Beyond their role in shaping the SCL, CTLs also modulate solute drag behavior. Neutral, singly ionized and doubly ionized acceptor dopants interact differently with the charged GB core, and therefore influence GB kinetics in distinct ways. In the following section, we analyze in detail how CTLs affect the strength of solute drag during GB migration.

\subsubsection{Charge transition level effects on solute drag strength}

To evaluate the influence of CTLs on solute drag and the resulting GB kinetics, the total driving force, which reflects the solute-drag strength during GB migration at prescribed velocities ($v_\text{GB}$), can be expressed as \cite{vikrant2020electrochemical}:
\be \label{FT}
\begin{aligned}
F_T &= F_{T,\eta} + F_{T,\VOne} + F_{T,\V} + F_{T,\AccZero} + F_{T,\AccOne} + F_{T,\AccTwo}\\
&=\int_{-\infty}^{\infty}\left[\frac{v_\text{GB}}{L}\left(\frac{\partial \eta}{\partial x}\right)^2 + \mu\subVOne^\text{ech}\left(\frac{\partial C\subVOne}{\partial x}\right) + \mu\subV^\text{ech}\left(\frac{\partial C\subV}{\partial x}\right) + \mu\subAccZero^\text{ech}\left(\frac{\partial C\subAccZero}{\partial x}\right) + \mu\subAccOne^\text{ech}\left(\frac{\partial C\subAccOne}{\partial x}\right) + \mu\subAccTwo^\text{ech}\left(\frac{\partial C\subAccTwo}{\partial x}\right)  \right] \text{d}x.
\end{aligned}
\ee
where the first term, $F_{T,\eta}$, accounts for the phase-field order parameter $\eta$, while the subsequent terms, $F_{T,\VOne}$, $F_{T,\V}$,  $F_{T,\AccZero}$, $F_{T,\AccOne}$, $F_{T,\AccTwo}$, represent the electrochemical contributions from oxygen vacancies and the different charge states of acceptor dopants. Because of their high mobilities, electrons and holes rapidly equilibrate across the GB region, and therefore their direct contribution to the total driving force are negligible. 

\begin{figure}[h]
    \centering \includegraphics[width=0.9\linewidth]{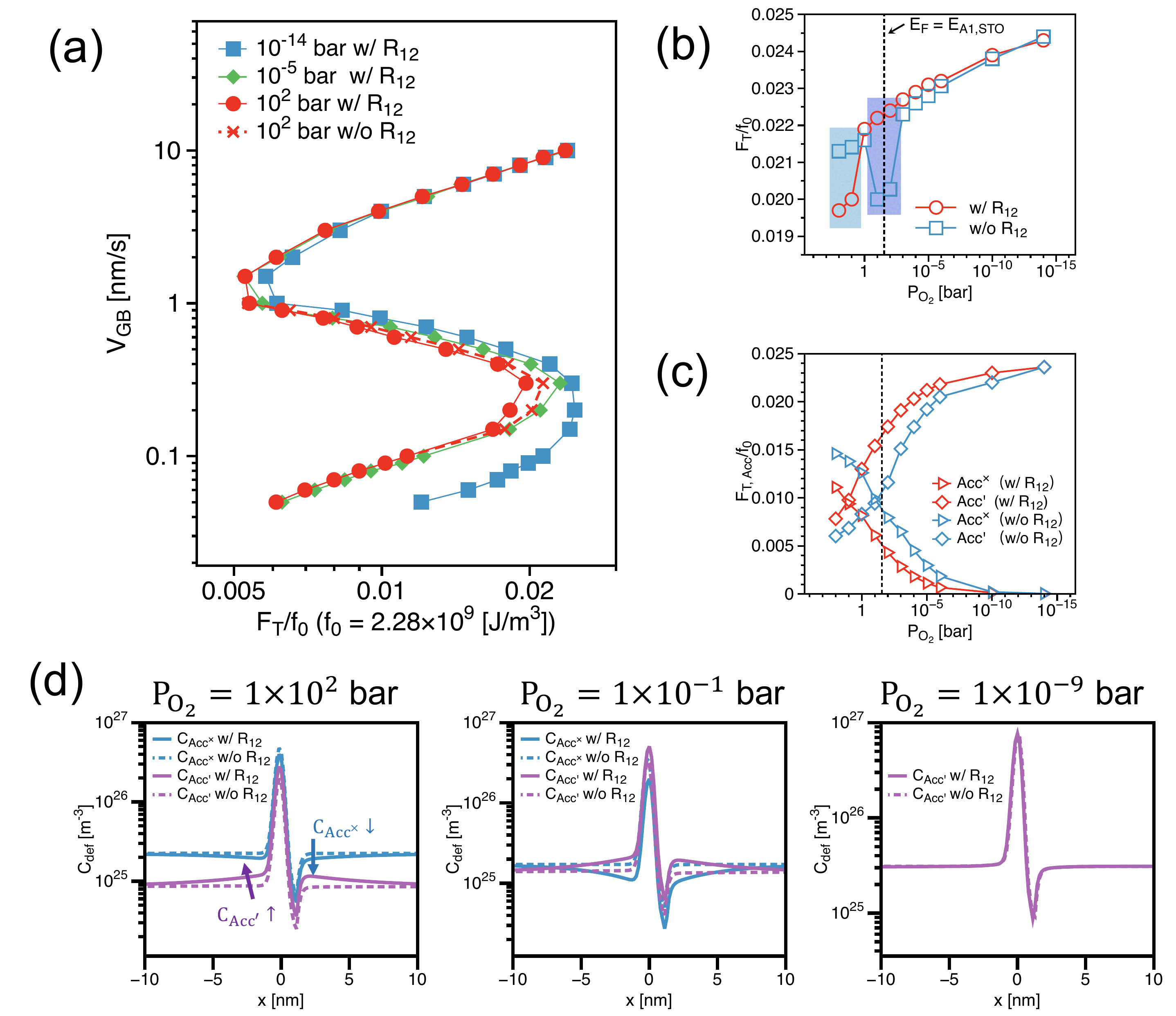}
    \caption{(a) GB velocity as a function of total driving force in Fe-doped STO under three representative oxygen partial pressures ($10^2$, $10^{-5}$, and $10^{-9}$~bar, corresponding to regimes 1, 2, and 3) at 1623~K. To assess the role of ultra-fast hole transport associated with ionization reactions $R_\text{ion1}$ and $R_\text{ion2}$, simulations were also performed with these reactions suppressed (denoted w/o $R_{12}$) and compared to the CTL-consistent case (w/ $R_{12}$). (b) Comparison of the total driving force at the inflection velocity (0.3~nm/s) for different oxygen partial pressures, highlighting the opposite trend in solute drag strength when the Fermi level is below ($10^{2}$~bar) or near ($10^{-1}$~bar) the Fe$^{4+/3+}$ CTL, and the convergence of both cases when the Fermi level approaches the conduction-band minimum ($10^{-14}$~bar). (c) Contributions of neutral and singly ionized acceptor dopants to the total driving force in w/ and w/o $R_{12}$ cases, showing that hole transport reduces the neutral contribution while enhancing that of the singly ionized species. (d) Spatial distributions of neutral and singly ionized acceptor dopants across the SCL for w/ and w/o $R_{12}$ cases. In the CTL-consistent case, hole transport increases the concentration of singly ionized dopants while reducing that of neutral dopants under $10^{2}$ and $10^{-1}$~bar. In contrast, under $10^{-9}$~bar, singly ionized dopants dominate, and CTLs have a negligible influence on both their spatial distribution and the resulting solute drag behavior.}
    \label{solute_drag_force}
\end{figure}

The numerical integration of \Cref{FT} under different oxygen partial pressures is shown in \Cref{solute_drag_force} (a). Three representative conditions, e.g. $10^{-14}$, $10^{-5}$, and $10^{2}$~bar, are selected to exemplify regime 1, 2, and 3, respectively. The corresponding total driving forces for GB velocities ranging from 0.05~nm/s to 10~nm/s are presented in \Cref{solute_drag_force}(a). The influence of CTL and Fermi level on GB kinetics are illustrated.
Additionally, for comparison, we also simulate a hypothetical scenario where ionization reactions are suppressed, i.e., $\nu\subAccZero = \nu\subAccOne = \nu\subAccTwo = 0$ in \Cref{AccOne_Con_eq}–\Cref{AccZero_Con_eq}, denoted as w/o $R_{12}$ in \Cref{solute_drag_force}. This choice effectively eliminates the ultra-fast electronic exchange processes (hole and electron transfer) that normally mediate the charge-state transitions of acceptor dopants, thereby allowing us to disentangle the contribution of pure dopant diffusion within the SCL. Although such a case does not occur under realistic conditions, it provides a useful reference to quantify the contribution of fast electronic exchange on solute drag.

We now turn to the total driving force extracted from the phase-field simulations. As shown in \Cref{solute_drag_force}(a), the relationship between GB velocity and driving force exhibits a characteristic S-shaped dependence and similar inflections for different oxygen partial pressures. For velocities below approximately 0.3~nm/s, the system corresponds to a slow-GB regime, in which the acceptor dopants migrate together with the GB. At velocities above 1~nm/s, a fast-GB regime emerges, where the GB breaks away from the dopant cloud. The intermediate range between 0.3 and 1~nm/s represents a transition regime, in which partial coupling between the dopants and the GB persists.

In the fast-GB regime ($v_\mathrm{GB} > 1$~nm/s), the total driving force is governed almost entirely by $F_{T, \eta}$. As a result, the values obtained at different oxygen partial pressures are nearly identical and exhibit a linear dependence on GB velocity, since the dopant distribution cannot respond to the rapid boundary motion. In contrast, in the slow-GB regime ($v_\mathrm{GB} < 0.3$~nm/s), the driving force arises predominantly from the sum of the multivalent acceptor contributions, i.e. $F_{T,\AccZero} + F_{T,\AccOne} + F_{T,\AccTwo}$, and thus shows a pronounced dependence on oxygen partial pressure. Under oxidizing condition (e.g. $10^{2}$~bar, regime~1), neutral acceptor dopants dominate, and the attraction between the positively charged GB and neutral dopants is weak, leading to a reduced solute drag effect. As the oxygen partial pressure decreases, singly ionized acceptors become increasingly dominant, and the solute drag effect is correspondingly enhanced (see $10^{-5}$~bar, regime~2). Under strongly reducing conditions (e.g., $10^{-14}$~bar, regime~3), the concentration of singly ionized acceptors greatly exceeds that of neutral dopants, and the solute drag contribution reaches a maximum.  This behavior is consistent with Cahn’s classical solute drag model \cite{cahn1962impurity}, in which the driving force contributed by acceptor dopants at low velocities can be approximated as $F_{T, \text{Acc}} \approx [\frac{1}{L} + \alpha C\subAcc(\infty)] v_\text{GB}$ in low velocity regime, with $\alpha \propto \int_{\infty}^{-\infty} \left\{\left[\sinh^2\frac{\Delta \mu\subAcc + z_i e \phi(x)}{2\bc T}\right]/D\subAcc \right\}\text{d}x$. For a positively charged GB ($\phi(x)>0$), the electrostatic potential across the SCL increases the coefficient $\alpha$ for negatively charged acceptors, thereby amplifying the solute drag effect. In contrast, for neutral acceptors, only the segregation energy contributes, without additional electrostatic enhancement.
These results highlight that the strength of solute drag is highly sensitive to the dominant charge state of acceptor dopants guided by CTLs.

In addition to the CTL-governed equilibrium charge state of acceptor dopants, the ultra-fast transport of electrons and holes dynamically re-equilibrates their charge-state distribution across the SCL, thereby exerting a significant influence on the solute drag effect. To isolate this contribution, phase-field simulations were performed at $10^{2}$~bar with ionization reactions $\eqref{Rion1}$ and $\eqref{Rion2}$ completely suppressed (denoted as w/o $R_{12}$, see red cross symbols with dashed line in \Cref{solute_drag_force}(a)). Results in the fast-GB regime are not shown, as they are essentially identical to the CTL-consistent cases. In the slow-GB regime, however, the comparison clearly demonstrates that ultra-fast hole transport reduces the effective solute drag strength by continuously re-equilibrating the dopant charge states across the SCL.

To provide further insight, additional simulations are performed at the inflection velocity (approximately 0.3~nm/s) with and without ionization reactions under different oxygen partial pressures, denoted as w/ $R_{12}$ and w/o $R_{12}$, respectively. The resulting total driving force as a function of oxygen partial pressure is shown in \Cref{solute_drag_force}(b) for both cases, while the separate solute drag contributions from neutral and singly ionized acceptor dopants are illustrated in \Cref{solute_drag_force}(c). The corresponding concentration distributions of neutral and singly ionized acceptor dopants across SCL are presented in \Cref{solute_drag_force}(d) for different oxygen partial pressures. 

As shown in \Cref{solute_drag_force}(b), under oxidizing conditions (e.g., $10^2$ and $10$~bar), the total solute drag strength in the case with suppressed holes transport and ionization reactions (w/o $R_{12}$) is higher than in the full CTL-consistent case (w/ $R_{12}$). When the oxygen partial pressure decreases to moderately reducing levels (e.g., $10^{-1}$ and $10^{-2}$~bar), the situation is reversed. The solute drag strength in the CTL-consistent case (w/ $R_{12}$) exceeds that of the w/o $R_{12}$ case. Under strongly reducing conditions, the two cases yield comparable solute drag strengths. These results clearly demonstrate that ultra-fast hole transport has a significant and oxygen-pressure-dependent impact on the solute drag effect.

The influence of ultra-fast hole transport on solute drag strength can be understood by analyzing the charge-state distribution of acceptor dopants in the SCL [\Cref{solute_drag_force}(d)]. At low velocities, solute drag in Fe-doped STO originates from both neutral and singly ionized acceptor dopants [\Cref{solute_drag_force}(c)]. 
Under oxidizing conditions (e.g., $10^2$~bar), the Fermi level lies below the Fe$^{4+/3+}$ CTL, and neutral acceptor dopants dominate, contributing much more strongly than singly ionized acceptor dopants. In the CTL-consistent case (w/ $R_{12}$), downward bending of the Fe$^{4+/3+}$ CTL at the positively charged GB core promotes hole-mediated conversion of neutral to singly ionized acceptor dopants. As a result, the neutral acceptor dopants decrease while the singly ionized acceptor dopants increase. Because neutral acceptor dopants remain the majority species and provide the larger contribution to solute drag [\Cref{solute_drag_force}(c)], their reduction dictates the overall response, leading to a lower solute-drag strength compared to the w/o $R_{12}$ case.
As the oxygen partial pressure decreases to moderately reducing levels (e.g., $10^{-1}$~bar), the Fermi level shifts above the Fe$^{4+/3+}$ CTL, and singly ionized acceptor dopants outnumber neutral acceptor dopants. The downward bending of the CTL further increases the concentration of singly ionized acceptor dopants in the SCL, thereby enhancing their contribution to solute drag. Moreover, the negative charge of singly ionized acceptor dopants couples strongly with the positive GB potential, amplifying their interaction and further strengthening the drag effect. Consequently, under such conditions, the CTL-consistent case exhibits a higher solute-drag strength than the w/o $R_{12}$ case.
Under strongly reducing conditions (e.g., $10^{-9}$~bar), the Fermi level lies far above the Fe$^{4+/3+}$ CTL and close to the conduction-band minimum. In this case, singly ionized acceptor dopants greatly outnumber neutral and doubly ionized acceptor dopants. In this regime, bending of the Fe$^{4+/3+}$ CTL induces only minor changes in the distribution of singly ionized acceptor dopants within the SCL, and consequently exerts a negligible effect on the overall solute drag strength.

Therefore, the dominant charge state of acceptor dopants determined by CTLs, together with charge-state transitions within the SCL induced by CTL bending, can significantly influence solute drag effects during GB migration. This study mainly focuses on defect-chemistry scenario with Fe$^{4+/3+}$ specific to Fe-doped STO. More generally, the influence of CTLs on solute drag is expected to be system-dependent and may vary considerably in materials with different defect-chemistry parameters. Two aspects are particularly noteworthy. On the one hand, variations in doping level, dopant species, GB type, oxygen partial pressure, and temperature can alter the dominant charge state of dopants, the associated SCL formation, and the GB potential, thereby modifying the extent of CTL bending. On the other hand, while  the Fe$^{4+/3+}$ CTL is mainly considered here, other dopants such as Mn exhibit different CTL energetics (e.g., Mn$^{4+/3+}$ and Mn$^{3+/2+}$) in BTO, with the Mn$^{3+/2+}$ CTL positioned lower in energy than that of Fe$^{3+/2+}$, which could result in different solute drag behaviors.

\subsection{Role of charge transition level and solute drag in shaping thermo-history–dependent grain boundary properties}

One important application of the present phase-field model is to predict GB properties that depend simultaneously on thermal history and GB type. The distribution of acceptor dopants established at high temperature during sintering influences the subsequent SCL formation during electrical measurements at lower temperature, where the redistribution of these dopants becomes frozen.

In this part, we focus on how thermo-history–dependent SCLs, shaped by CTLs and solute drag effects during sintering, govern GB properties such as the GB potential, space-charge width, and GB resistance in bicrystals. 
To address this, we assume that samples are sintered and equilibrated at 1623~K and $P_{\mathrm{O}_2} = 10^{-5}$~bar under different GB migration velocities and subsequently quenched rapidly to the measurement temperature (e.g., 600~K) and $P_{\mathrm{O}_2} = 10^{-5}$~bar. {Upon quenching, the acceptor dopants retain their quasi-equilibrium concentration profiles established under different GB velocities during sintering. In addition, the total concentration of oxygen vacancies is assumed to be frozen. This assumption is justified by the fact that, at temperatures below approximately 700~K, oxidation and reduction reactions at the solid–gas interface become kinetically limited. As a result, reaction \Cref{Rred} can be considered effectively frozen, and the total oxygen vacancy concentration remains constant during the subsequent defect redistribution and SCL reformation \cite{wang2016defect, usler2024space}.
In contrast, charge-state redistribution among oxygen vacancies and multivalent acceptor dopants remains active at the measurement temperature through reactions \Cref{Rred1}, \Cref{Rion1}, and \Cref{Rion2}.}

\begin{figure}[]
    \centering \includegraphics[width=0.9\linewidth]{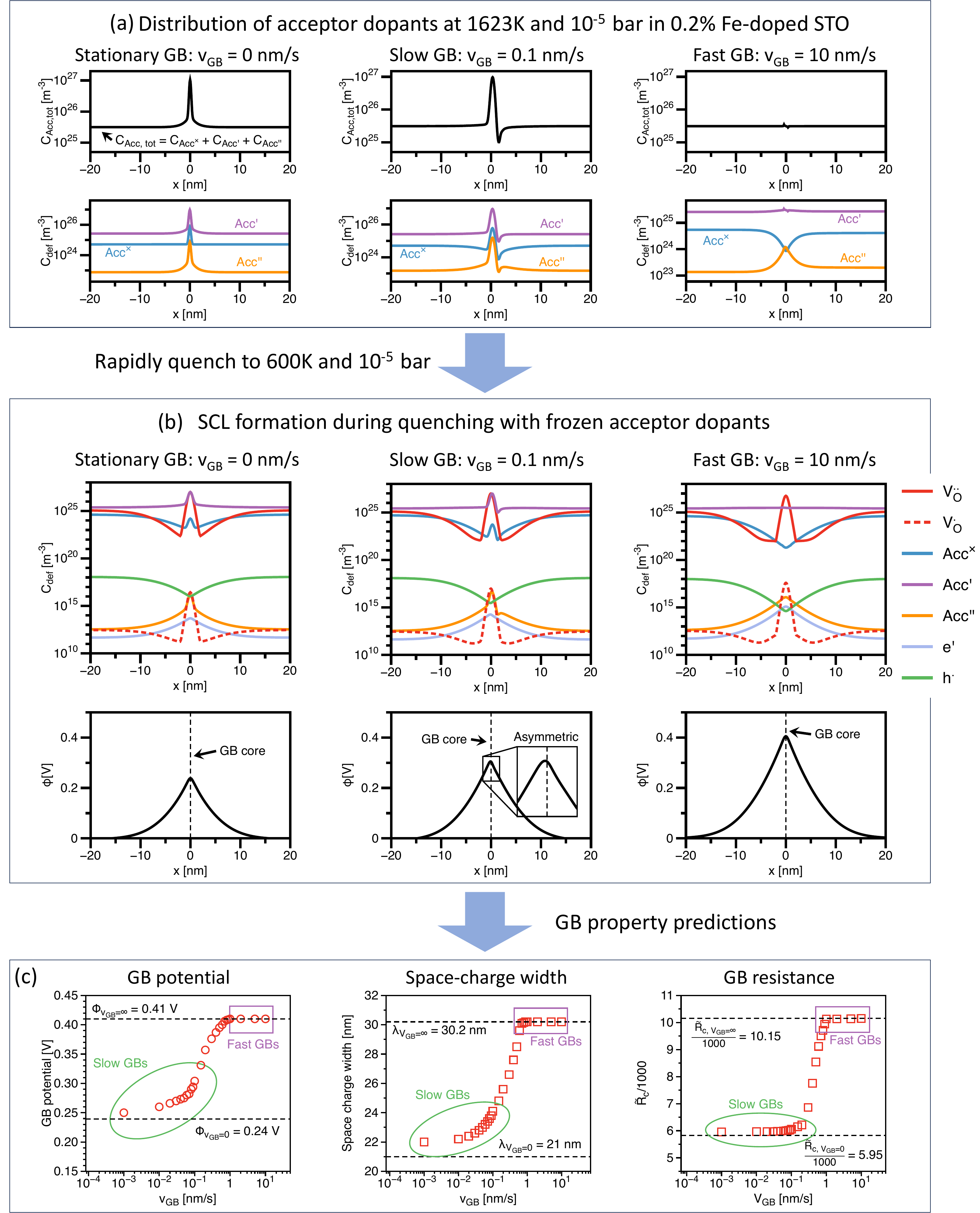}
    \caption{The phase-field model is employed to predict GB properties as a function of thermal history and GB type. Sintering is assumed at 1623~K and $P_{\mathrm{O}_2}=10^{-5}$~bar, while electrical measurements are conducted at 600K under the same oxygen partial pressure, where singly ionized acceptor dopants and doubly ionized oxygen vacancies dominate.
(a) Acceptor dopant concentration profiles for different GB velocities during sintering.
(b) Resulting SCL structures after rapid quenching to 600~K at $P_{\mathrm{O}_2}=10^{-5}$~bar. During quenching, the total acceptor dopant concentration is locally preserved, and the solid–gas reaction \Cref{Rred} is kinetically hindered.
(c) Predicted GB properties, including GB potential, space-charge width, and GB resistance.}
    \label{GB_property}
\end{figure}

The thermo-history–dependent SCLs formed at the measurement temperature and the resulting GB properties are presented in \Cref{GB_property}. Specifically, the total concentration of acceptor dopants established under different GB migration velocities during sintering at 1623~K and $P_{\mathrm{O}_2} = 10^{-5}$~bar is extracted from the phase-field simulations. After rapid quenching to 600~K at $P_{\mathrm{O}_2} = 10^{-5}$~bar, this total concentration of acceptor dopants is assumed to be frozen-in. At the same time, the distribution among neutral, singly ionized, and doubly ionized states is re-equilibrated through CTL-governed hole transport at 600~K, and this redistributed profile is taken as the initial condition in the phase-field simulations to obtain the final SCL profile at the measurement temperature. In contrast to the frozen-in dopants, oxygen vacancies, electrons, and holes can reach thermodynamic equilibrium at 600~K. The thermo-history–dependent SCL is then characterized by the GB potential and space-charge width, while the corresponding electrical response is quantified by the GB resistance.

These conditions are chosen only as a representative case for illustration. In practice, the specific sintering and measurement conditions need to be adjusted to reflect the actual experimental scenario. It should be emphasized that our aim is not to quantitatively predict the electrical properties of sintered samples at the measurement temperature. In practice, the thermal history of ceramics is more complex than the idealized rapid-quenching scenario assumed here. In particular, the cooling rate strongly influences the extent to which dopant distributions and associated space charge profiles are frozen-in. Slower cooling may allow partial re-equilibration above the restricted-equilibrium temperature before redistribution by diffusion process becomes fully suppressed at low temperatures.
Instead, our focus is to elucidate how fast and slow GBs, induced by solute drag effects during sintering, influence SCL formation and thereby affect the electrical response under the consideration of CTLs at low temperature. This approach is particularly significant as it not only captures the role of CTLs across different temperatures but also extends beyond conventional defect-chemistry calculations by explicitly incorporating moving GBs and the impact of solute drag on electrical performance.

Then, we briefly outline the procedure for calculating the GB resistance. The electrical resistance associated with a single GB in a bicrystal is treated as an excess quantity, defined as the deviation from the homogeneous bulk behavior. {When the contribution of oxygen vacancies dominates the total conductivity}, the GB resistance, $R_\text{GB}$, can be computed by integrating the inverse of the local total conductivity, denoted as $\sigma_\text{tot}(x)$ across the SCL region, and subtracting the corresponding bulk contribution \cite{maier2023physical}:
\be
R_\text{GB} = R_\text{tot} - R_\text{b} =  \frac{2}{A}\left(\int_{0}^{L_x} \frac{1}{\sigma_\text{tot}(x)} \text{dx} - \int_{0}^{L_x} \frac{1}{\sigma_\text{tot, b}(\infty)} \text{dx}\right),
\ee
where $A$ is the cross-sectional area of the grain. $L_\text{x}$ is the half-width of one grain and is much larger compared with the space-charge width. $\sigma_\text{tot, b}(\infty)$ is the bulk total conductivity far from the GB core, while $\sigma_\text{tot}(x)$ is the spatially dependent total conductivity including the contributions of doubly ionized oxygen vacancies, electrons, and holes. Due to very low singly ionized oxygen vacancy concentration, its contribution to the total conductivity is negligible. Then, the total conductivity expression is given by
\be
\sigma_\text{tot}(x) = \sigma\subV(x) + \sigma\subE(x) + \sigma\subHole(x),
\ee
where $\sigma\subV(x)$, $\sigma\subE(x)$ and $\sigma\subHole(x)$ are the local conductivities of oxygen vacancies, electrons and holes, respectively. They can be obtained via
\be
\sigma\subV(x) = |z\subV|u\subV e C\subVb h_\text{b},
\ee
\be
\sigma\subE(x) = |z\subE|u\subE e C\subE h_\text{b},
\ee
\be
\sigma\subHole(x) = |z\subHole|u\subHole e C\subHole h_\text{b},
\ee
where $u\subV$, $u\subE$ and $u\subHole$ are the temperature dependent mobilities of oxygen vacancies, electrons and holes, respectively. The estimated mobilities are listed in \Cref{ParameterTable}. {Assuming that the resistance of the GB core is negligible \cite{tong2020analyzing}, the conductivities of oxygen vacancies, electrons and holes are evaluated only in the bulk phase.} $C\subVb$ is the partitioning concentrations of oxygen vacancy in bulk, which can be obtained from phase-field simulations. $h_\text{b}$ is the interpolation functions (see \Cref{Free energy density functional}). The local concentrations of electrons and holes, $C\subE$ and $C\subHole$, are evaluated from Eqs.~\eqref{Ce_EF} and \eqref{Ch_EF}, with $C\subE = C\subEb = C\subEc$ and $C\subHole = C\subHoleb = C\subHolec$ (see \Cref{evolution_equations}).
Here we define a dimensionless resistance, denoted as $\tilde{R}_\text{GB}$,  as the ratio of the GB resistance to the bulk resistance:
\be
\tilde{R}_\text{GB} = \frac{R_\text{GB}}{R_\text{b}} = \int_{0}^{L_x} \left[\frac{\sigma_\text{tot, b}(\infty)}{\sigma_\text{tot}(x)} -1 \right]\text{dx}. 
\ee

In \Cref{GB_property}(a), three representative acceptor dopant distributions at 1623~K are obtained for different GB velocities: 0~nm/s for a stationary GB, 0.1~nm/s for a slow GB, and 10~nm/s for a fast GB. 
The Fermi level is 1.17~eV, which is beyond the Fe$^{4+/3+}$ CTL (0.94~eV). The distributions of neutral, singly ionized, doubly ionized, and total dopants are shown. For a stationary GB, dopants segregate at the GB core and form symmetric accumulation zones within the SCL. For a slow GB, solute drag effect strongly influences GB migration, breaking the symmetry of the SCL and transforming accumulation zone into depletion zone. For a fast GB, the dopants cannot keep pace with the migrating GB, and their distribution becomes nearly uniform within the SCL.


The re-equilibrated distributions of acceptor dopants are then used as the initial conditions for the phase-field simulations, from which the equilibrium SCLs are obtained at 600~K and $10^{-5}$~bar. The resulting SCL formation is analyzed for three representative cases as demonstrated in \Cref{GB_property}(b): a stationary GB (without solute drag), a slow GB, and a fast GB, the latter two reflecting different regimes of solute drag.  For the stationary GB, the SCL remains symmetric. Singly and doubly ionized acceptor dopants continue to segregate at the GB core, compensating the positively charged oxygen vacancies. Meanwhile, a depletion zone of neutral acceptor dopants develops, and the neutral-to-singly ionized charge-state transition becomes significant within the SCL due to the Fe$^{4+/3+}$ bending.  For the slow GB, the asymmetric distribution of acceptor dopants formed at high temperature gives rise to an asymmetric SCL. The concentration of acceptor dopants on the left side of the SCL is higher than on the right side, resulting in distinct compensation scenarios across the SCL.
For the fast GB, the distribution of singly ionized acceptor dopants within the SCL becomes nearly uniform, and the resulting profile closely resembles the idealized behavior predicted by the classical Mott–Schottky model.

During sintering, the microstructure consists of a mixture of slow and fast GBs, and the GB velocities are distributed over a continuous range rather than restricted to discrete values. Consequently, the GB properties measured at low temperature should not be regarded as single values but as distributed quantities that reflect this velocity spectrum. These properties are bounded by two limiting cases: the lower bound corresponds to the stationary limit ($v_\text{GB}=0$), and the upper bound corresponds to the case of infinitely fast boundaries ($v_\text{GB}\to\infty$), where dopants remain uniformly distributed, identical to the assumption of the classical Mott–Schottky model.

In \Cref{GB_property}(c), the GB properties, including the GB potential, space-charge width, and GB resistance, are predicted as functions of GB velocity at the measurement temperature.
For the GB potential, fast GBs approach the value of 0.41V predicted by the Mott–Schottky model, while slow GBs exhibit reduced values. In the stationary limit, the GB potential is significantly lower, with the GB potential ratio between stationary and fast GBs being approximately 0.585. For moving GBs, the ratio between slow and fast GBs increases with GB velocity, from about 0.61 at 0.001~nm/s to 0.92 at 0.3~nm/s.
A similar trend is observed for the space-charge width. Fast GBs exhibit the largest width of approximately 30.2~nm, whereas slow GBs show progressively narrower widths. The width of slow GBs corresponds to about 0.728–0.881 of that of fast GBs for moving boundaries (from 0.001~nm/s to 0.3~nm/s), and decreases further to approximately 0.695 in the stationary limit.
For GB resistance, it converges toward this upper bound as the GB velocity increases. The dimensionless value $\tilde{R}_\text{c}$ reaches a maximum of about 10150. This result is close to the defect-chemistry prediction based on
$\tilde{R}_\text{c} = \frac{l_\text{D}}{L_\text{x}}\frac{\exp(2e\Phi_0/k_\mathrm{B}T)}{\sqrt{4e\Phi_0/k_\mathrm{B}T}} = 9626$,
where $l_\text{D}=2.87~\text{nm}$ is the Debye length, $\Phi_0=0.41~\text{V}$ is the GB potential and $L_\text{x}=408~\text{nm}$ is half-width of one grain \cite{tong2020analyzing}. 
By contrast, when the GB velocity decreases, the resistance drops rapidly, and most of the slow GB cases collapse toward the stationary limit ($v_\text{GB}=0$). The GB resistance of slow GBs corresponds to 0.586 of that of fast GBs. 

Experimental measurement of the properties of GBs in undoped STO polycrystalline was carried out by Zahler et. al \cite{zahler2025non}. They observed two types GB (GB$_1$ and GB$_2$) and the GB potential, space-charge width and the GB resistance ratios between GB$_1$ and GB$_2$ are 0.62, 0.77 and 0.38. 
It should be noted that the present work considers Fe-doped STO with only one GB, whereas the experimental study by Zahler et al. was performed on undoped STO and the different GBs may interact. Despite this difference, an interesting qualitative correspondence can be found. In Zahler’s work, the GB potential, space-charge width, and GB resistance of GB$_1$ were all smaller than those of GB$_2$. In our simulations, the properties of slow GBs are consistently smaller than those of fast GBs. Furthermore, the experimentally measured GB potential and space-charge width ratios fall well within the range predicted by our simulations. 

In addition, the Mott–Schottky model was employed in Zahler’s work to extract SCL parameters from EIS measurements for two types of GBs \cite{zahler2025non}. Our simulations indicate that this assumption is only valid for fast GBs, where the distribution of acceptor dopants remains nearly uniform and their influence on SCL formation at the measurement temperature is negligible. Under such conditions, the SCL profile closely resembles the idealized Mott–Schottky prediction. In contrast, for slow GBs the solute drag effect together with CTLs leads to a spatially inhomogeneous distribution of acceptor dopants, which in turn strongly modifies the local charge balance and electrostatic potential. As a result, the Mott–Schottky approximation becomes inadequate, and a more detailed description that explicitly accounts for dopant segregation and charge-state transitions is required to capture the actual SCL formation.

\section{Conclusions}

In this work, we have developed a defect-chemistry consistent phase-field model that explicitly incorporates the Fermi level and CTLs into the description of SCL formation at stationary and migrating GBs. Going beyond established phase-field models that fix dopant charge states, our model accounts for the charge-state transitions of multivalent dopants in the bulk and, importantly, within SCLs, where bending of CTLs drives local changes in dopant valence. 
The main results of this paper include:
\begin{itemize}
\item The effects of Fe$^{4+/3+}$ and Fe$^{3+/2+}$ CTLs on bulk defect chemistry were examined by calculating defect concentrations in 0.2\% Fe-doped STO at 600~K and 1623~K, revealing four distinct regimes. At high oxygen partial pressures, neutral acceptor dopants are predominant (Regime~1). At intermediate oxygen partial pressures, singly ionized acceptor dopants dominate and are compensated by doubly ionized oxygen vacancies (Regime~2). At very low oxygen partial pressures, oxygen vacancies and electrons become the majority species (Regime 3). 

\item  With consideration of CTL-influenced bulk defect chemistry, we simulated the formation of symmetric SCLs at stationary GBs in 0.2\% Fe-doped STO at 600~K and 1623~K under different oxygen partial pressures. The GB potential varies from 0.07~V to 0.4~V at 600~K, whereas it ranges from -0.075~V to 0.15~V at 1623~K under different oxygen partial pressures. At 600~K, although acceptor dopants are frozen-in, their local charge states can still be redistributed within the SCL as a consequence of CTL bending. In contrast, at 1623~K, where acceptor dopants are mobile, both dopant diffusion and CTL bending jointly govern SCL formation, resulting in a fully equilibrated defect distribution. 

\item  During GB migration at 1623~K, our simulations reveal the formation of asymmetric SCLs induced by solute drag in 0.2\% Fe-doped STO under different oxygen partial pressures. The degree of asymmetry increases with GB velocity. In particular, at intermediate oxygen partial pressures (regime~2), the SCL in front of the fast GB extends significantly, introducing an enlarged charge-state transition zone. Moreover, the GB potential rises with GB velocity, thereby enhancing charge-state transitions within the SCL. 
 
\item The influence of CTLs on solute drag strength was also examined. In the fast GB regime, CTLs exert negligible influence because the acceptor dopants remain uniformly distributed within the SCL owing to their very low diffusivity. In contrast, within the slow GB regime, neutral acceptor dopants may dominate the SCL under high oxygen partial pressures, leading to a pronounced reduction in solute drag strength. As the oxygen partial pressure decreases, the fraction of singly ionized acceptor dopants increases, which in turn enhances the solute drag strength. 
Beyond the diffusion-controlled redistribution of acceptor dopants, the ultra-fast transport of holes caused by the ionization reactions in the CTL bending region further facilitates charge-state transitions of multivalent acceptor dopants, providing an additional pathway through which CTLs regulate the solute drag strength. 

\item We applied this phase-field model to predict GB properties that depend simultaneously on thermal history and GB type. The concentration profiles of acceptor dopants established at the sintering temperature for different GB velocities were retained and subsequently redistributed under electrical measurement conditions. The resulting equilibrium SCLs reveal clear distinctions between slow and fast GBs: slow GBs exhibit smaller GB potential, narrower space-charge width, and lower GB resistance than fast GBs. Furthermore, the simulations suggest that the commonly used Mott–Schottky assumption is appropriate for fast GBs when extracting space-charge characteristics from impedance spectroscopy data, whereas for slow GBs this assumption may cause discrepancies due to the non-uniform distribution of acceptor dopants.
\end{itemize}

Overall, this work establishes a unified framework that links defect chemistry, Fermi level, CTLs, and space-charge behavior at GBs. While the present study focuses on bicrystal configurations, the GPU-accelerated phase-field framework can be readily extended to three-dimensional polycrystalline systems to capture microstructure evolution during sintering. This approach opens new opportunities for predictive modeling of doped perovskite oxides and provides guidance for the design of functional ceramics with tailored GB properties.

\section{Acknowledgment}
The financial support of German Science Foundation (DFG) in the framework of the Collaborative Research Centre 1548 (CRC 1548, project number 463184206) and the Project 471260201 are acknowledged. The authors K. Wang and B.-X. Xu greatly appreciate the access to the Lichtenberg II High-Performance Computer (HPC) and the technique supports from the HHLR, Technische
Universität Darmstadt. The computing time on the HPC is granted by the NHR4CES Resource
Allocation Board under the project “special00007”.

\appendix

\section{Electrochemical potentials of different defects in phase-field model}
The electrochemical potentials of all defect species are derived from the free energy.
Their explicit expressions are 

\begin{equation}
\mu\subVOneb^ \text{ech} = \frac{\partial f^\text{ech}_\text{b}}{\partial C\subVOneb}=\mu^0_\text{V,b} + \bc T \ln\left(\frac{C\subVOneb}{N_\text{V,b} - C\subVb - C\subVOneb}\right) + z\subVOne e \phi,
\end{equation}
\begin{equation}
\mu\subVOnec^ \text{ech} = \frac{\partial f^\text{ech}_\text{c}}{\partial C\subVOnec}=\mu^0_\text{V,c} + \bc T \ln\left(\frac{C\subVOnec}{N_\text{V,c} - C\subVc - C\subVOnec}\right) + z\subVOne e \phi,
\end{equation}

\begin{equation}
\mu\subVb^ \text{ech} = \frac{\partial f^\text{ech}_\text{b}}{\partial C\subVb}=\mu^0_\text{V,b} + \bc T \ln\left(\frac{C\subVb}{N_\text{V,b} - C\subVb - C\subVOneb}\right) + z\subV e \phi,
\end{equation}
\begin{equation}
\mu\subVc^ \text{ech} = \frac{\partial f^\text{ech}_\text{c}}{\partial C\subVc}=\mu^0_\text{V,c} + \bc T \ln\left(\frac{C\subVc}{N_\text{V,c} - C\subVc - C\subVOnec}\right) + z\subV e \phi,
\end{equation}

\begin{equation}
\mu\subAccOneb^ \text{ech} = \frac{\partial f^\text{ech}_\text{b}}{\partial C\subAccOneb}=\mu^0\subAccb + \bc T \ln\left(\frac{C\subAccOneb}{N\subAccb - C\subAccOneb - C\subAccTwob - C\subAccZerob}\right) + z\subAccOne e \phi,
\end{equation}
\begin{equation}
\mu\subAccOnec^ \text{ech} = \frac{\partial f^\text{ech}_\text{c}}{\partial C\subAccOnec}=\mu^0\subAccc + \bc T \ln\left(\frac{C\subAccOnec}{N\subAccc - C\subAccOnec - C\subAccTwoc - C\subAccZeroc}\right) + z\subAccOne e \phi,
\end{equation}

\begin{equation}
\mu\subAccTwob^ \text{ech} = \frac{\partial f^\text{ech}_\text{b}}{\partial C\subAccTwob}=\mu^0\subAccb + \bc T \ln\left(\frac{C\subAccTwob}{N\subAccb - C\subAccOneb - C\subAccTwob - C\subAccZerob}\right) + z\subAccTwo e \phi,
\end{equation}
\begin{equation}
\mu\subAccTwoc^ \text{ech} = \frac{\partial f^\text{ech}_\text{c}}{\partial C\subAccTwoc}=\mu^0\subAccc + \bc T \ln\left(\frac{C\subAccTwoc}{N\subAccc - C\subAccOnec - C\subAccTwoc - C\subAccZeroc}\right) + z\subAccTwo e \phi,
\end{equation}

\begin{equation}
\mu\subAccZerob^ \text{ech} = \frac{\partial f^\text{ech}_\text{b}}{\partial C\subAccZerob}=\mu^0\subAccb + \bc T \ln\left(\frac{C\subAccZerob}{N\subAccb - C\subAccOneb - C\subAccTwob - C\subAccZerob}\right) + z\subAccZero e \phi,
\end{equation}
\begin{equation}
\mu\subAccZeroc^ \text{ech} = \frac{\partial f^\text{ech}_\text{c}}{\partial C\subAccZeroc}=\mu^0\subAccc + \bc T \ln\left(\frac{C\subAccZeroc}{N\subAccc - C\subAccOnec - C\subAccTwoc - C\subAccZeroc}\right) + z\subAccZero e \phi,
\end{equation}

\begin{equation}\label{muEb}
\mu\subEb^ \text{ech} = \frac{\partial f^\text{ech}_\text{b}}{\partial C\subEb}=\mu^0\subEb + \bc T \ln\left(\frac{C\subEb}{N_\text{CB,b} - C\subEb}\right) + z\subE e \phi,
\end{equation}
\begin{equation}\label{muEc}
\mu\subEc^ \text{ech} = \frac{\partial f^\text{ech}_\text{c}}{\partial C\subEc}=\mu^0\subEc + \bc T \ln\left(\frac{C\subEc}{N_\text{CB,c} - C\subEc}\right) + z\subE e \phi,
\end{equation}

\begin{equation}\label{muHb}
\mu\subHoleb^ \text{ech} = \frac{\partial f^\text{ech}_\text{b}}{\partial C\subHoleb}=\mu^0\subHoleb + \bc T \ln\left(\frac{C\subHoleb}{N_\text{VB,b} - C\subHoleb}\right) + z\subHole e \phi,
\end{equation}
\begin{equation}\label{muHc}
\mu\subHolec^ \text{ech} = \frac{\partial f^\text{ech}_\text{c}}{\partial C\subHolec}=\mu^0\subHolec + \bc T \ln\left(\frac{C\subHolec}{N_\text{VB,c} - C\subHolec}\right) + z\subHole e \phi.
\end{equation}
The electrochemical free energies and electrochemical potentials presented here represent the general cases applicable to both bulk and GB core regions, consistent with defect chemistry theory \cite{maier2023physical}.

\bibliographystyle{elsarticle-num} 
\bibliography{reference}

\end{document}